\documentclass[12pt, a4paper]{article}
\usepackage{lipsum}
\usepackage{amsmath}
\usepackage{amssymb}
\usepackage{amsthm}
\usepackage{mathtools}
\usepackage{fancyhdr}
\usepackage{graphicx}
\usepackage{verbatim}
\usepackage{latexsym}
\usepackage{mathrsfs}
\usepackage{mathdots}

\usepackage{tikz}
\usepackage{tikz-cd}

\usepackage{luatex85} 
\usepackage[all]{xy}

\usepackage{float}
\usepackage{authblk}
\usepackage{siunitx}
\usepackage{eqparbox}

\usepackage{todonotes}
\setuptodonotes{inline}

\usepackage[shortlabels]{enumitem}
\setlist[itemize]{leftmargin=*}

\usepackage{pdfpages}
\usepackage{xcolor}
\usepackage[calcwidth]{titlesec}
\usepackage{stmaryrd}
\usepackage{wallpaper}
\usepackage{setspace}
\usepackage[top = 2 cm, bottom = 3 cm, left = 2 cm, right = 2 cm]{geometry}
\usepackage{bbm}

\usepackage[UKenglish]{isodate}
\cleanlookdateon

\usepackage{anyfontsize}
\usepackage{pgf,pgfplots}
\usepackage{textcomp}
\usepackage[margin = 1cm]{caption}
\usepackage{stackengine}

\usepackage{subcaption}

\usepackage{wasysym}
\usepackage{esint} 

\usepackage[style=alphabetic,
  sorting = nyt,
  sortcites=true,
  giveninits=true,
  date=year,
  isbn=false,
  maxbibnames=99,
  maxalphanames=6,
  backend=biber]{biblatex}
\DeclareFieldFormat*{titlecase}{\MakeSentenceCase{#1}}

\DeclareSourcemap{\maps[datatype=bibtex]{\map{\step[fieldset=shorthand, null]}}}

\DeclareSourcemap{
  \maps[datatype=bibtex]{
    \map[overwrite]{
      \step[fieldsource=doi, final]
      \step[fieldset=url, null]
      \step[fieldset=eprint, null]
    }  
  }
}

\DeclareSourcemap{
  \maps[datatype=bibtex, overwrite]{
    \map{
      \step[fieldset=language, null]
      \step[fieldset=month, null]
      \step[fieldset=urldate, null]
    }
  }
}

\AtEveryBibitem{%
  \clearlist{language}%
  \clearlist{month}%
  \clearlist{urldate}%
  \clearfield{urldate}%
  \ifentrytype{inproceedings}
    {\clearlist{booktitle}%
     \clearlist{publisher}%
     \clearlist{location}}
}

\renewbibmacro{in:}{}

\DeclareFieldFormat[misc]{title}{\mkbibquote{#1}}
\DeclareFieldFormat[article,inproceedings]{volume}{\mkbibbold{#1}}
\DeclareFieldFormat[inproceedings]{series}{\mkbibitalic{#1}}

\usepackage[hidelinks,breaklinks=true]{hyperref}
\usepackage[capitalise,noabbrev]{cleveref}

\usepackage{bookmark}

\usepackage{seqsplit}
\usepackage[notcite,notref,final]{showkeys}

\pgfplotsset{compat=1.16}
\usetikzlibrary{arrows}
\usetikzlibrary{arrows.meta}
\usetikzlibrary{patterns}
\usetikzlibrary{calc}
\usetikzlibrary{math} 

\usetikzlibrary{positioning,decorations.pathreplacing} 

\usepackage{chngcntr}
\counterwithin{figure}{section}
\counterwithin{equation}{section}

\let\C\relax

\newcommand{\R}{\mathbb{R}}
\newcommand{\C}{\mathbb{C}}

\newcommand{\N}{\mathbb{N}}
\newcommand{\Z}{\mathbb{Z}}

\newcommand{\mcS}{\mathcal{S}}
\newcommand{\mcG}{\mathcal{G}}
\newcommand{\mcC}{\mathcal{C}}
\newcommand{\mcT}{\mathcal{T}}
\newcommand{\mcL}{\mathcal{L}}
\newcommand{\mcA}{\mathcal{A}}
\newcommand{\mcD}{\mathcal{D}}
\newcommand{\mcN}{\mathcal{N}}
\newcommand{\mcB}{\mathcal{B}}
\newcommand{\mcV}{\mathcal{V}}
\newcommand{\mcW}{\mathcal{W}}
\newcommand{\mcM}{\mathcal{M}}
\newcommand{\mcH}{\mathcal{H}}
\newcommand{\mcZ}{\mathcal{Z}}
\newcommand{\mcR}{\mathcal{R}}

\newtheoremstyle{theorems}
  {3pt}
  {3pt}
  {\itshape}
  {}
  {\bfseries}
  {.}
  { }
  {}

\newtheoremstyle{proofparts}
  {3pt}
  {0pt}
  {}
  {\parindent}
  {\scshape}
  {:}
  {\newline}
  {}

\newtheoremstyle{claims}
  {2pt}
  {2pt}
  {}
  {\parindent}
  {\bfseries}
  {.}
  { }
  {}

\makeatletter
\newcommand{\newreptheorem}[2]{\newtheorem*{rep@#1}{\rep@title}\newenvironment{rep#1}[1]{\def\rep@title{#2 \ref*{##1}}\begin{rep@#1}}{\end{rep@#1}}}
\makeatother

\theoremstyle{theorems}
\newtheorem{thm}{Theorem}[section]

\newtheorem{lemma}[thm]{Lemma}
\newtheorem*{lemma*}{Lemma}

\newreptheorem{thm}{Theorem}
\newreptheorem{lemma}{Lemma}

\theoremstyle{definition}
\newtheorem{defn}[thm]{Definition}

\newtheorem{remark}[thm]{Remark}

\theoremstyle{proofparts}

\makeatletter
\@addtoreset{proofpart}{thm}
\makeatother

\theoremstyle{claims}

\newtheorem*{claim*}{Claim}

\crefname{thm}{theorem}{theorems}
\crefname{problem}{problem}{problems}
\crefname{lemma}{lemma}{lemmas}
\crefname{cor}{corollary}{corollaries}
\crefname{prop}{proposition}{propositions}
\crefname{conj}{conjecture}{conjectures}
\crefname{defn}{definition}{definitions}
\crefname{note}{note}{notes}
\crefname{ex}{example}{examples}
\crefname{remark}{remark}{remarks}
\crefname{notation}{notation}{notations}
\crefname{assumption}{assumption}{assumptions}
\crefname{claim}{claim}{claims}
\crefname{claim*}{claim}{claims}

\renewcommand{\subsectionmark}[1]{}
\makeatletter
\newcommand{\Biggg}{\bBigg@{3}}
\newcommand{\vast}{\bBigg@{4}}
\newcommand{\Vast}{\bBigg@{5}}
\makeatother

\newcommand{\norm}[1]{\left\Vert #1 \right\Vert}
\newcommand{\abs}[1]{\left\vert #1 \right\vert}

\DeclareMathOperator{\Vol}{Vol}

\definecolor{emphcolor}{rgb}{0,0,1}           

\newcommand{\ip}[2]{\left\langle #1 \middle\vert #2 \right\rangle}
\newcommand{\longip}[3]{\left\langle #1 \middle\vert #2 \middle\vert #3 \right\rangle}

\newcommand{\ud}{\,\textnormal{d}}

\let\epsilon\varepsilon
\let\varepsilon\epsilon

\newcommand{\eps}{\epsilon}

\allowdisplaybreaks

\title{Almost optimal upper bound for the ground state energy of a dilute Fermi gas via cluster expansion}
\author{Asbjørn Bækgaard Lauritsen\thanks{\href{mailto:alaurits@ist.ac.at}{\nolinkurl{alaurits@ist.ac.at}}}}
\affil{Institute of Science and Technology Austria, Am Campus 1, 3400 Klosterneuburg, Austria}

\begin{document}
\maketitle

\begin{abstract}
We prove an upper bound on the energy density of the dilute spin-$\frac{1}{2}$ Fermi gas
capturing the leading correction to the kinetic energy $8\pi a \rho_\uparrow\rho_\downarrow$
with an error of size smaller than $a\rho^{2}(a^3\rho)^{1/3-\eps}$ for any $\eps > 0$,
where $a$ denotes the scattering length of the interaction.
The result is valid for a large class of interactions including interactions with a hard core.
A central ingredient in the proof is a rigorous version of a fermionic cluster expansion adapted from 
the formal expansion of Gaudin, Gillespie and Ripka (Nucl. Phys. A, 176.2 (1971), pp. 237--260).
\end{abstract}

\section{Introduction and main results}
We consider an interacting Fermi gas of $N$ particles
interacting via a two-body interaction $v$ which we assume to be non-negative, radial and of compact support.
In units where $\hbar = 1$ and the particle mass is $m=1/2$ the Hamiltonian is given by 
\[
H_N = \sum_{j=1}^N - \Delta_j + \sum_{i < j} v(x_i - x_j),
\]
where $\Delta_j$ denotes the Laplacian on the $j$'th coordinate.
For spin-$\frac{1}{2}$ fermions in some domain $\Lambda = \Lambda_L = [-L/2, L/2]^3$
one realises the Hamiltonian on the space 
$L^2_a(\Lambda^{N}, \C^2) = \bigwedge^N L^2(\Lambda, \C^2)$.
Since the Hamiltonian is spin-independent we can specify 
definite values for the number of particles with each spin,
i.e. $N_\sigma$ particles of spin $\sigma \in \{\uparrow,\downarrow\}$
and $N_\uparrow + N_\downarrow = N$.
In this setting the Hamiltonian is realized on the space 
$\mcH_{N_\uparrow,N_\downarrow} = L^2_a(\Lambda^{N_\uparrow}) \otimes L^2_a(\Lambda^{N_\downarrow})$.
The ground state energy on the space $L^2_a(\Lambda^{N}, \C^2)$
is then given by minimizing 
in $N_\sigma$ (satisfying $N_\uparrow + N_\downarrow = N$)
the ground state energies on the spaces $\mcH_{N_\uparrow,N_\downarrow}$.

This system was previously studied in \cite{Lieb.Seiringer.ea.2005,Falconi.Giacomelli.ea.2021,Giacomelli.2023} where it is shown that 
for a dilute system in the thermodynamic limit
\[
e(\rho_\uparrow, \rho_\downarrow)
=
\lim_{\substack{L\to \infty \\ N_\sigma/L^3 \to \rho_\sigma}}
  \inf_{\substack{\psi \in \mcH_{N_\uparrow,N_\downarrow} \\ \norm{\psi}_{L^2} = 1}} 
    \frac{\longip{\psi}{H_N}{\psi}}{L^3}
=
\frac{3}{5} (6\pi^2)^{2/3} (\rho_\uparrow^{5/3} + \rho_\downarrow^{5/3})
+ 8\pi a \rho_\uparrow\rho_\downarrow 
+ a\rho^2 \eps(a^3\rho),
\]
where $\rho = \rho_\uparrow + \rho_\downarrow$, 
$a$ is the \emph{($s$-wave) scattering length} of the interaction $v$
and $\eps(a^3\rho) = o(1)$ in the limit $a^3\rho \ll 1$.
The existence of the thermodynamic limit follows from \cite{Robinson.1971}.
Moreover the limit doesn't depend on the boundary conditions.

The leading term $\frac{3}{5} (6\pi^2)^{2/3} (\rho_\uparrow^{5/3} + \rho_{\downarrow}^{5/3})$ 
is the kinetic energy of a free Fermi gas. 
The next term $8\pi a \rho_\uparrow\rho_\downarrow$ is the leading correction coming from the interaction.
This term may be understood as coming from the energy of a pair of opposite-spin fermions times the number of such pairs.
The energy correction arising from interactions between fermions of the same spin is of order $a_p^3 \rho^{8/3}$,
where $a_p$ denotes the $p$-\emph{wave scattering length} (see \cite{Lauritsen.Seiringer.2024})
and so much smaller.

The first proof of this result was given by 
Lieb, Seiringer and Solovej \cite{Lieb.Seiringer.ea.2005}.
Their proof gives the explicit error bounds 
$-C (a^3\rho)^{1/39} \leq \eps(a^3\rho) \leq C (a^3\rho)^{2/27}$
for some constant $C > 0$.
These error bounds were later improved in \cite{Falconi.Giacomelli.ea.2021} and very recently in \cite{Giacomelli.2023}, 
where in particular the ``optimal'' upper bound $\eps(a^3\rho)\leq C (a^3\rho)^{1/3}$ is shown.
The works \cite{Falconi.Giacomelli.ea.2021,Giacomelli.2023}
however deal with more regular potentials than
the quite general setting studied in \cite{Lieb.Seiringer.ea.2005},
where it is assumed that the interaction is non-negative, radial and compactly supported.
In \cite{Falconi.Giacomelli.ea.2021,Giacomelli.2023}
the interaction is additionally assumed to be smooth.
In particular interactions with a hard core are not treated in \cite{Falconi.Giacomelli.ea.2021,Giacomelli.2023}.

The upper bound of order $a\rho^{1/3}$ is optimal in the sense 
that this is the order of the conjectured next term in the expansion. 
Namely the Huang--Yang term \cite{Huang.Yang.1957}, see \cite{Giacomelli.2023a,Giacomelli.2023}.

Our main theorem is the ``almost optimal'' upper bound $\eps(a^3\rho) \leq C_\delta (a^3\rho)^{1/3 - \delta}$
for any $\delta > 0$ for some $\delta$-dependent constant $C_\delta > 0$
under the same assumptions as in \cite{Lieb.Seiringer.ea.2005}, i.e. weaker than that of \cite{Falconi.Giacomelli.ea.2021,Giacomelli.2023}.
In particular we allow for $v$ to have a hard core.
A central ingredient in the proof 
is to prove a rigorous version of a fermionic cluster expansion adapted from \cite{Gaudin.Gillespie.ea.1971}.
This is analogous to what is done in \cite{Lauritsen.Seiringer.2024} for spin-polarized fermions.
(See also \cite{Lauritsen.Seiringer.2023a} for the application to spin-polarized fermions at positive-temperature.)

\subsection{Precise statements of results}
To give the statement of our main theorem, we first define the 
\emph{scattering length(s)} of the interaction $v$.
\begin{defn}[{\cites[Appendix A]{Lieb.Yngvason.2001}[Section 4]{Seiringer.Yngvason.2020}}]
\label{def.scattering.length}
The $s$- and $p$-\emph{wave scattering lengths} $a$ and $a_p$ are defined by 
\[
\begin{aligned}
4\pi a 
& = \inf 
\left\{\int \left(|\nabla f|^2 + \frac{1}{2}v|f|^2\right) \ud x
: f(x) \to 1 \textnormal{ for } |x|\to \infty
\right\},
\\
12\pi a_p^3 
& = \inf 
\left\{\int \left(|\nabla f|^2 + \frac{1}{2}v|f|^2\right) |x|^2 \ud x
: f(x) \to 1 \textnormal{ for } |x|\to \infty
\right\}.
\end{aligned}
\]
The minimizing $f$'s are the $s$- and $p$-\emph{wave scattering functions}.
They are denoted $f_{s0}$ and $f_{p0}$ respectively.
\end{defn}
\noindent
The minimizing functions $f_{s0}$ and $f_{p0}$ are real-valued.
We collect properties of them in \Cref{lem.properties.scattering.function}.

With this we may then state our main theorem.
\begin{thm}\label{thm.main}
Let $0\leq v\leq +\infty$ be radial and of compact support. Then 
for any $\delta > 0$ and for sufficiently small $a^3\rho$ we have 
\[
e(\rho_\uparrow,\rho_\downarrow) \leq 
\frac{3}{5} (6\pi^2)^{2/3} (\rho_\uparrow^{5/3} + \rho_\downarrow^{5/3})
+ 8\pi a \rho_\uparrow\rho_\downarrow 
+ O_\delta\left(a\rho^{2}(a^{3}\rho)^{1/3 - \delta}\right).
\]
\end{thm}

\noindent
The subscript $\delta$ in $O_\delta$ denotes that the implicit constant depends on $\delta$.
Further, the $v$-dependence of the error-term $O_\delta\left(a\rho^{2}(a^{3}\rho)^{1/3 - \delta}\right)$ 
is only via the scattering lengths $a$ and $a_p$
(meaning that the implicit constant depends on the ratio $a_p/a$ but otherwise not on $v$).
In particular we note that $v$ is allowed to have a hard core, meaning $v(x) = +\infty$ for $|x|\leq r_0$ for some $r_0 > 0$.

The essential steps of the proof are 
\begin{enumerate}[(1)]
\item 
Show the absolute convergence of a fermionic cluster expansion adapted from the formal calculations of \cite{Gaudin.Gillespie.ea.1971}.
For this we need the ``Fermi polyhedron'', a polyhedral approximation to the Fermi ball, described in \cite[Section 2.2]{Lauritsen.Seiringer.2024}.
The calculation of the fermionic cluster expansion is given in \Cref{sec.gaudin} and the absolute convergence in shown in \Cref{sec.abs.conv}.

\item 
Bound the energy of a Jastrow-type trial state. 
For this step, the central part is computing the values of all diagrams of a certain type exactly
and using these exact values up to some arbitrary high order.
This is somewhat similar to the approach in \cite{Basti.Cenatiempo.ea.2023a} for the dilute Bose gas.
This calculation is part of the proof of \Cref{lem.bound.error.rho.(1.1)}.
\end{enumerate}

\begin{remark}[Higher spin]
With not much difficulty one can extend the result to higher spin and with a spin-dependent interaction $v_{\sigma\sigma'} = v_{\sigma'\sigma}$.
The result for $S\geq 2$ spin values $\{1,\ldots, S\}$ is
\[
    e\left(\rho_1, \ldots, \rho_S\right)
    \leq 
    \frac{3}{5}(6\pi^2)^{2/3} 
    \sum_{\sigma=1}^S \rho_\sigma^{5/3}
    + 8\pi \sum_{1\leq \sigma < \sigma' \leq S} a_{\sigma \sigma'} \rho_\sigma \rho_{\sigma'}
    + O_\delta\left(a\rho^2(a^{3}\rho)^{1/3 - \delta}\right),
\]
where $a_{\sigma\sigma'}$ is the $s$-wave scattering length 
of the spin $\sigma$ -- spin $\sigma'$ interaction $v_{\sigma\sigma'}$
and $a = \sup_{\sigma < \sigma'} a_{\sigma \sigma'}$.
For conciseness of the proof we will only give it for $S = 2$, i.e. for spin-$\frac{1}{2}$ fermions.
We will however give comments on how to adapt the 
individual (non-trivial) steps of the proof to the higher spin setting.
These comments are given in \Cref{rmk.higher.spin.(1.1.1),rmk.higher.spin.diagrams.def,rmk.higher.spin.multinomial,rmk.higher.spin.no.diagrams}.
\end{remark}

\begin{remark}[{Comparison with \cite{Falconi.Giacomelli.ea.2021,Lieb.Seiringer.ea.2005,Giacomelli.2023}}]\label{rmk.compare.method}
The trial state we consider, $\psi_{N_\uparrow,N_\downarrow}$ (defined in \Cref{eqn.trial.state} below), 
is in spirit the same as that considered in \cite{Lieb.Seiringer.ea.2005}.
They differ only in technical aspects (discussed in \Cref{rmk.compare.trial.state} below).
The reason we are able to improve on the bound in \cite{Lieb.Seiringer.ea.2005} is that we treat the 
cancellations between $\longip{\psi}{H_N}{\psi}$ and $\ip{\psi}{\psi}$ 
more precisely
(for the [non-normalized] trial state $\psi$ being defined as $\psi_{N_\uparrow,N_\downarrow}$ in \Cref{eqn.trial.state} 
only without the normalization constant $C_{N_\uparrow,N_\downarrow}$).

In \cite{Falconi.Giacomelli.ea.2021,Giacomelli.2023} a completely different method is employed. 
There the system is studied using a method inspired by Bogoliubov theory for dilute Bose gases. 
(The ``bosons'' appearing here as pairs of opposite-spin fermions.)
\end{remark}

\noindent
The paper is structured as follows.
In \Cref{sec.preliminary} we give some preliminary computations and 
recall some properties of the scattering functions and Fermi polyhedron from \cite{Lieb.Yngvason.2001,Lauritsen.Seiringer.2024}.
Next, in \Cref{sec.gaudin} we give the calculation of a 
fermionic cluster expansion adapted from \cite{Gaudin.Gillespie.ea.1971}.
Subsequently, in \Cref{sec.abs.conv} we find conditions for the absolute convergence 
of the cluster expansion formulas of \Cref{sec.gaudin}.
Finally, in \Cref{sec.energy} we use the formulas of \Cref{sec.gaudin} to bound the 
energy of a Jastrow-type trial state.

\section{Preliminary computations}\label{sec.preliminary}
We first give a few preliminary computations. 
We will construct a trial state using a box method of glueing trial states in smaller boxes together in \Cref{sec.box}.
In the smaller boxes we will need to use Dirichlet boundary conditions, however in \Cref{sec.box}
we will construct trial states with Dirichlet boundary conditions out of trial states with periodic boundary conditions.
(See also \cite{Lauritsen.Seiringer.2024}.)
Thus, we will use periodic boundary conditions in the box $\Lambda = \Lambda_L = [-L/2, L/2]^3$.

First we establish some notation.

\subsection{Notation}
\begin{itemize}
\item 
We write $x_i$ and $y_j$ for the spatial coordinates of particle $i$ of spin $\uparrow$ respectively particle $j$ of spin $\downarrow$.

We write $z_i$ to mean either $x_i$ or $y_i$ if the spin is not important.

We write additionally $z_{(i,\uparrow)} = x_i$ and $z_{(i,\downarrow)} = y_i$.

\item 
We write
$[n,m] = \{n,n+1,\ldots,m\}$ for integers $n\leq m$.
If $n > m$ then $[n,m] = \varnothing$.

\item
For a set $A$ we write $Z_A = (z_a)_{a\in A}$ for the coordinates of the 
vertices with labels in $A$. (Similarly for $X_A$ and $Y_A$.)

In particular we write $Z_{[n,m]} = (z_n,\ldots,z_m)$ for the coordinates of 
particles $n,n+1,\ldots,m$.

If $n=1$ we simply write $Z_m = Z_{[1,m]} = (z_1,\ldots,z_m)$.

\item 
We write $C$ for a generic (positive) constant, whose value may change line by line.
If we want to emphasize the dependence on some parameter $A$ we will denote this by $C_A$.
\end{itemize}

\noindent
We consider the indices of the coordinates as vertices $\mu=(i,\sigma)\in V_{\infty,\infty}:= \N\times \{\uparrow,\downarrow\}$.
Here $\sigma \in \{\uparrow,\downarrow\}$ labels the spin of the particle.
Then we define
\[
  V_{n,m} 
  := V_{n}^\uparrow \cup V_{m}^\downarrow,
  \qquad 
  V_p^\sigma := \{(1,\sigma),\ldots,(p,\sigma)\} \subset V_{\infty,\infty}, 
  \quad \sigma \in \{\uparrow,\downarrow\},
  \quad p \in \N \cup \{\infty\}.
\]
(We mean $V_\infty^\sigma = \N \times \{\sigma\}$ for $p=\infty$.)
On the vertices $V_{\infty,\infty}$ we define the ordering $<$ as follows.
\[
  \mu = (i,\sigma) < (j,\sigma') = \nu
  \qquad 
  \overset{\textnormal{def}}{\Longleftrightarrow}
  \qquad 
  \left(\sigma=\uparrow \textnormal{ and } \sigma'=\downarrow \right)
  \textnormal{ or }
  \left(\sigma=\sigma' \textnormal{ and } i < j\right).
\]
Define the rescaled and cut-off scattering functions $f_s$ and $f_p$ as 
\begin{equation}\label{eqn.define.fs.fp}
f_s(x) = \begin{cases}
    \frac{1}{1 - a/b} f_{s0}(|x|) & |x|\leq b,
    \\
    1 & |x|\geq b,
\end{cases}
\qquad 
f_p(x) = \begin{cases}
    \frac{1}{1 - a_p^3/b^3} f_{p0}(|x|) & |x|\leq b,
    \\
    1 & |x|\geq b,
\end{cases}
\end{equation}
where $|\cdot| := \inf_{n\in \Z^3}|\cdot - nL|_{\R^3}$ (with $|\cdot|_{\R^3}$ denoting the norm on $\R^3$),
$b = \rho^{-1/3}$
and the scattering function $f_{s0}$ and $f_{p0}$ are defined in \Cref{def.scattering.length}.
(They are radial functions, see \Cref{lem.properties.scattering.function}, so $f_{s}$ and $f_p$ are well-defined.) 
We prefer to write $b$ instead of its value $\rho^{-1/3}$ to keep apparent dependences on $b$.
For $b=\rho^{-1/3}$ we have $b > R_0$, the range of $v$, for sufficiently small $a^3\rho$. 
Hence, $f_{s}$ and $f_p$ are continuous for sufficiently small $a^3\rho$.
(Note that the metric on the torus $\Lambda$ is given by $d(x,y) = |x-y|$.
We will abuse notation slightly and denote by $|\cdot|$ also the absolute value of some number or the norm on $\R^3$.)

To simplify notation we write for $\mu,\nu \in V_{\infty,\infty}$
\begin{equation}\label{eqn.def.f.mu.nu}
  f_{\mu\nu} := \begin{cases}
  f_p(x_i - x_j) & \mu = (i,\uparrow), \nu = (j,\uparrow),
  \\
  f_s(x_i - y_j) & \mu = (i,\uparrow), \nu = (j,\downarrow),
  \\
  f_s(y_i - x_j) & \mu = (i,\downarrow), \nu = (j,\uparrow),
  \\
  f_p(y_i - y_j) & \mu = (i,\downarrow), \nu = (j,\downarrow),
  \end{cases}
\end{equation}
and similar for all quantities derived from $f_{s}$ and $f_p$.
In particular $\nabla f_{\mu\nu} = \nabla f_{s/p}(z_\mu - z_\nu)$ with $s/p$ meaning $s$ if the spins of $\mu$ and $\nu$ are different 
and $p$ if they are the same.

Next, we introduce the (non-normalized) Slater determinants $D_{N_\uparrow}$ and $D_{N_\downarrow}$ as
\[
D_{N_\sigma}(Z_{N_\sigma}) 
= \det \left[ u_{k}(z_i)\right]_{\substack{k\in P_F^\sigma \\ i = 1,\ldots, N_\sigma}},
\qquad 
N_\sigma = \# P_F^\sigma,
\qquad 
u_k(z) = L^{-3/2} e^{ikz},
\]
where $P_F^\sigma$ is the ``Fermi polyhedron'', a polyhedral approximation to the Fermi ball described in \Cref{sec.fermi.polyhedron}, 
see also \cite[Section 2.2]{Lauritsen.Seiringer.2024},
and $\# P_F^\sigma$ denotes the number of points in $P_F^\sigma$.

Further, we denote for $\mu,\nu \in V_{\infty,\infty}$
\begin{equation}\label{eqn.def.gamma.mu.nu}
\gamma_{\mu\nu} := 
  \begin{cases}
  \gamma_{N_\uparrow}^{(1)}(x_i; x_j) & \mu = (i,\uparrow), \nu = (j,\uparrow),
  \\
  0 & \mu = (i,\uparrow), \nu = (j,\downarrow),
  \\
  0 & \mu = (i,\downarrow), \nu = (j,\uparrow),
  \\
  \gamma_{N_\downarrow}^{(1)}(y_i; y_j) & \mu = (i,\downarrow), \nu = (j,\downarrow),
  \end{cases}
\end{equation}
where $\gamma_{N_\sigma}^{(1)}$ are the one-particle density matrices of $\frac{1}{\sqrt{N_\uparrow!}} D_{N_\uparrow}$ and $\frac{1}{\sqrt{N_\downarrow!}}D_{N_\downarrow}$.

Finally, for any (normalized) state $\psi\in L^2_a(\Lambda^{N_\uparrow}) \otimes L^2_a(\Lambda^{N_\downarrow})$
we will normalize reduced densities as follows (for $n+m\geq 1$). 
\begin{equation}\label{eqn.normalize.rho.nm}
\begin{aligned}
\rho_{\psi}^{(n,m)}
 & = N_\uparrow(N_\uparrow - 1) \cdots (N_\uparrow - n + 1)
    N_\downarrow(N_\downarrow - 1)\cdots (N_\downarrow - m + 1)
\\ & \quad \times 
    \idotsint \abs{\psi}^2
    \ud X_{[n+1,N_\uparrow]} \ud Y_{[m+1, N_\downarrow]}.
\end{aligned}   
\end{equation}
For a (normalized) Slater determinant 
$\psi = \psi(X_{N_\uparrow},Y_{N_\downarrow}) = \frac{1}{\sqrt{N_\uparrow! N_\downarrow!}} D_{N_\uparrow}(X_{N_\uparrow}) D_{N_\downarrow}(Y_{N_\downarrow})$
we write $\rho^{(n,m)} = \rho_{\psi}^{(n,m)}$
and for the trial state $\psi_{N_\uparrow,N_\downarrow}$ we write $\rho^{(n,m)}_{\textnormal{Jas}} = \rho_{\psi_{N_\uparrow,N_\downarrow}}^{(n,m)}$.

We will fix the Fermi momenta $k_F^\sigma$ such that the ratio $k_F^\uparrow / k_F^\downarrow$ is rational, see \Cref{rmk.k_F.rational}. 
This is a restriction on which densities $\rho_\sigma$ can arise from the trial state $\psi_{N_\uparrow,N_\downarrow}$, see \Cref{rmk.dependent.parameters}.
We extend to all densities in \Cref{sec.box}.
The dilute limit will be realized as $(k_F^\uparrow + k_F^\downarrow) a \to 0$.

\subsection{Computation of the energy}
We consider the trial state 
\begin{multline}\label{eqn.trial.state}
  \psi_{N_\uparrow,{N_\downarrow}}(X_{N_\uparrow},Y_{N_\downarrow}) 
    = \frac{1}{\sqrt{C_{N_\uparrow,{N_\downarrow}}}} 
  \left[\prod_{\substack{\mu,\nu \in V_{N_\uparrow,N_\downarrow} \\ \mu < \nu}} f_{\mu\nu}\right]
    D_{N_\uparrow}(X_{N_\uparrow})
    D_{N_\downarrow}(Y_{N_\downarrow})
  \\
    = \frac{1}{\sqrt{C_{N_\uparrow,N_\downarrow}}} 
    \left[\prod_{\substack{1\leq i \leq N_\uparrow \\ 1 \leq j \leq {N_\downarrow}}} f_s(x_i - y_j)
            \prod_{1\leq i < j \leq N_\uparrow} f_p(x_{i} -  x_{j}) 
            \prod_{1\leq i < j \leq {N_\downarrow}} f_p(y_{{i}} - y_{j})\right]
    D_{N_\uparrow}(X_{N_\uparrow})
    D_{N_\downarrow}(Y_{N_\downarrow}),
\end{multline}
where $C_{N_\uparrow,N_\downarrow}$ is a normalization constant such that 
$\int \abs{\psi_{N_\uparrow,N_\downarrow}}^2 \ud X_{N_\uparrow} \ud Y_{N_\downarrow} = 1$.

\begin{remark}[{Comparison to \cite{Lieb.Seiringer.ea.2005}}]\label{rmk.compare.trial.state}
As mentioned in \Cref{rmk.compare.method} the trial state $\psi_{N_\uparrow,N_\downarrow}$ is mostly the same as that of \cite{Lieb.Seiringer.ea.2005}. 
They differ in two technical aspects: 
\begin{enumerate}[1.]
\item The choice of function implementing the correlations between particles of the same spin.

The exact function used is not particularly important 
since the same-spin interactions give rise to a much smaller energy correction (than that of different-spin interactions). 
The function $f_p$ is a natural choice.

\item The choice of Slater determinant.

Our choice of Slater determinants with momenta in the Fermi polyhedron (as opposed to the Fermi ball, which is what is used in \cite{Lieb.Seiringer.ea.2005}) 
is a technical necessity as we discuss in \Cref{sec.fermi.polyhedron} below. 
\end{enumerate}
\end{remark}

\noindent
We compute the energy of the trial state $\psi_{N_\uparrow,N_\downarrow}$
\[
\longip{\psi_{N_\uparrow,N_\downarrow}}{H_N}{\psi_{N_\uparrow,N_\downarrow}}
= 
\idotsint 
\left[
\sum_{\mu\in V_{N_\uparrow,N_\downarrow}}\abs{\nabla_{z_\mu} \psi_{N_\uparrow,N_\downarrow}}^2
+ \sum_{\substack{\mu,\nu \in V_{N_\uparrow,N_\downarrow} \\ \mu < \nu}} v(z_\mu - z_\nu) \abs{\psi_{N_\uparrow,N_\downarrow}}^2
\right]
\!\ud X_{N_\uparrow} \ud Y_{N_\downarrow}.
\]
Note that for (real-valued) functions $F,G$ we have 
\begin{equation}\label{eqn.int.nabla.FG}
\int \abs{\nabla (FG)}^2 
= -\int G\Delta G |F|^2 + \int |G|^2 \abs{\nabla F}^2.
\end{equation}
By symmetries of the Fermi polyhedron, see \Cref{def.P_F}, we have that $D_{N_\uparrow}$ and $D_{N_\downarrow}$ are real-valued.
Thus, using \Cref{eqn.int.nabla.FG} for 
$F = \prod_{\mu < \nu} f_{\mu\nu}$ and $G = D_{N_\uparrow}D_{N_\downarrow}$
for each of the derivatives $\nabla_{x_i}$, $\nabla_{y_j}$
we get (recall that $\nabla f_{\mu\nu} = \nabla f_{s/p}(z_\mu - z_\nu)$)
\begin{equation*}
\begin{aligned}
& 
\sum_{\mu \in V_{N_\uparrow,N_\downarrow}} \idotsint \abs{\nabla_{z_\mu} \psi_{N_\uparrow,N_\downarrow}}^2 \ud X_{N_\uparrow} \ud Y_{N_\downarrow}
\\ & \quad  = 
E_0^\uparrow + E_0^{\downarrow} +
\idotsint 
\ud X_{N_\uparrow} \ud Y_{N_\downarrow}
\abs{\psi_{N_\uparrow,N_\downarrow}}^2
\Biggg[
2 \sum_{\mu \in V^\uparrow_{N_\uparrow}} \sum_{\nu \in V^\downarrow_{N_\downarrow}} \abs{\frac{\nabla f_{\mu\nu}}{f_{\mu\nu}}}^2
+ 2 \sum_{\sigma \in \{\uparrow, \downarrow\}}
\sum_{\substack{\mu, \nu \in V^\sigma_{N_\sigma} \\ \mu < \nu}} \abs{\frac{\nabla f_{\mu\nu}}{f_{\mu\nu}}}^2
\\ & \qquad 
+ \sum_{\sigma \in \{\uparrow,\downarrow\}}
\sum_{\mu \in V^\sigma_{N_\sigma}} \sum_{\substack{\nu,\lambda \in V^{-\sigma}_{N_{-\sigma}} \\ \nu \ne \lambda}} 
\frac{\nabla f_{\mu\nu} \nabla f_{\mu\lambda}}{f_{\mu\nu} f_{\mu\lambda}} 
+ \sum_{\sigma \in \{\uparrow,\downarrow\}}
\sum_{\substack{\mu,\nu \in V^\sigma_{N_\sigma} \\ \mu \ne \nu}} \sum_{\lambda \in V^{-\sigma}_{N_{-\sigma}}} 
\frac{\nabla f_{\mu\nu} \nabla f_{\mu\lambda}}{f_{\mu\nu} f_{\mu\lambda}} 
\\ & \qquad 
- \sum_{\sigma \in \{\uparrow, \downarrow\}} 
\sum_{\substack{\mu,\nu,\lambda \in V_{N_\sigma}^\sigma \\ \mu, \nu,  \lambda \textnormal{ distinct}}}
\frac{\nabla f_{\mu\nu} \nabla f_{\nu\lambda}}{ f_{\mu\nu} f_{\nu\lambda}}
\Biggg],
\end{aligned}
\end{equation*}
where $E_0^\sigma = \sum_{k\in P_F^\sigma} |k|^2$ is the kinetic energy 
of the Slater determinants $\frac{1}{\sqrt{N_\sigma!}}D_{N_\sigma}$
and $-\sigma$ is the ``other spin'', i.e. $-\uparrow = \downarrow$ and $-\downarrow = \uparrow$.
(The factor $2$ in the term $2 \sum_{\mu \in V^\uparrow_{N_\uparrow}} \sum_{\nu \in V^\downarrow_{N_\downarrow}} \abs{\frac{\nabla f_{\mu\nu}}{f_{\mu\nu}}}^2$
arises as $2 = \sum_{\sigma \in \{\uparrow,\downarrow\}} 1$.)
The terms are grouped according to how many $s$-wave $f$'s appear.
In terms of the reduced densities we thus get 
\begin{equation}\label{eqn.energy.compute}
\begin{aligned}
& 
\longip{\psi_{N_\uparrow,N_\downarrow}}{H_N}{\psi_{N_\uparrow,N_\downarrow}}
\\ & \quad  = 
E_0^\uparrow + E_0^{\downarrow} +
2 
\iint 
\rho_{\textnormal{Jas}}^{(1,1)} 
\left[\abs{\frac{\nabla f_{s}(x_1-y_1)}{f_{s}(x_1-y_1)}}^2
+ \frac{1}{2}v(x_1-y_1)\right]
\ud x_1 \ud y_1 
\\ & \qquad 
+
  \iint \rho^{(2,0)}_{\textnormal{Jas}}
  \left[\abs{\frac{\nabla f_{p}(x_1-x_2)}{f_{p}(x_1-x_2)}}^2 
  + \frac{1}{2}v(x_1-x_2)\right]
  \ud x_1 \ud x_2 
\\ & \qquad 
  +
  \iiint \rho_{\textnormal{Jas}}^{(2,1)}
  \left[
  \frac{\nabla f_s(x_1 - y_1) \nabla f_s(x_2 - y_1)}{f_s(x_1-y_1) f_s(x_2-y_1)}
  + \frac{\nabla f_s(x_1-y_1) \nabla f_p(x_1-x_2)}{ f_s(x_1-y_1) f_p(x_1-x_2)}
  \right]
  \ud x_1 \ud x_2 \ud y_1
\\ & \qquad 
-
  \iiint \rho_{\textnormal{Jas}}^{(3,0)} \frac{\nabla f_p(x_1-x_2) \nabla f_p(x_2-x_3)}{f_p(x_1-x_2)f_p(x_2-x_3)} \ud x_1 \ud x_2 \ud x_3
  + \textnormal{terms with } \rho_{\textnormal{Jas}}^{(0,2)}, \rho_{\textnormal{Jas}}^{(1,2)}, \rho_{\textnormal{Jas}}^{(0,3)}.
\end{aligned}
\end{equation}
We find formulas for the reduced densities in \Cref{sec.gaudin}. Before doing so, 
we first recall some properties on the ``Fermi polyhedron'' $P_F^\sigma$ and the scattering functions $f_s, f_p$.

\subsection{Properties of the ``Fermi polyhedron'' and the scattering functions}\label{sec.fermi.polyhedron}
In this section we recall a few properties of the ``Fermi polyhedron'' from \cite[Section 2.2 and Lemma 4.9]{Lauritsen.Seiringer.2024} 
and scattering functions from \cite[Appendix A]{Lieb.Yngvason.2001}.

The reason for introducing the ``Fermi polyhedron'' is that for the analysis of the absolute convergence of the Gaudin--Gillespie--Ripka expansion 
we need good control over 
\begin{equation*}
\int_\Lambda \abs{\gamma_{N_\sigma}^{(1)}(x;0)} \ud x = \int_\Lambda \abs{\frac{1}{L^3}\sum_{k\in P_{F}^\sigma} e^{ikx}} \ud x.
\end{equation*}
By \Cref{eqn.derivative.lebesgue.constant.0} below (coming from \cite[Lemma 2.12]{Lauritsen.Seiringer.2024} and \cite{Kolomoitsev.Lomako.2018})
this is bounded by $s (\log N)^{3}$.
If we had instead chosen the Slater determinants in the trial state $\psi_{N_\uparrow,N_\downarrow}$ to have momenta in 
the Fermi ball $B_F^\sigma = \{ k\in \frac{2\pi}{L}\Z^3 : |k| \leq k_F^\sigma\}$ we would have \cite{Ganzburg.Liflyand.2019,Liflyand.2006}
\begin{equation*}
\int_\Lambda \abs{\gamma_{N_\sigma}^{(1)}(x;0)} \ud x = \int_\Lambda \abs{\frac{1}{L^3}\sum_{k\in B_{F}^\sigma} e^{ikx}} \ud x \sim N^{1/3}.
\end{equation*}
This $N$-dependence would prevent us from achieving that both the Gaudin--Gillespie--Ripka expansion converges absolutely
and that the finite-size error from the kinetic energy is negligible. See also \Cref{rmk.N.dependence.abs.conv} and \cite[Remark 3.5]{Lauritsen.Seiringer.2024}.

The ``Fermi polyhedron'' is defined in \cite[Definition 2.7]{Lauritsen.Seiringer.2024}. 
We give here only a sketch of the definition and state a few properties needed for our purposes.
For a full discussion with proofs we refer to \cite[Section 2.2 and Appendix B]{Lauritsen.Seiringer.2024}.

\begin{defn}[{Sketch, see 
\cite[Definition 2.7]{Lauritsen.Seiringer.2024}}]\label{def.P_F}
For each spin $\sigma\in \{\uparrow,\downarrow\}$ 
define the (convex) polyhedron $P^\sigma$ with $s_\sigma$ ``corners'' (extremal points) as follows.
All ``corners'' $\kappa_1^\sigma,\ldots,\kappa_{s_\sigma}^\sigma$ are chosen of the form 
$\kappa_j^\sigma = \zeta_\sigma (\frac{p^1_j}{Q_1^\sigma},\frac{p^2_j}{Q_2^\sigma},\frac{p^3_j}{Q_3^\sigma})$, where 
$\zeta_\sigma \in \R$, 
$p^i_j\in \Z$ for $i=1,2,3$, $j=1,\ldots,s_\sigma$ and $Q_1^\sigma,Q_2^\sigma,Q_3^\sigma$ are large distinct primes.
Then $P^\sigma$ is the convex hull of these ``corners'' and 
$\zeta_\sigma$ is chosen such that $\Vol P^\sigma = \frac{4\pi}{3}$.

The polyhedron $P^\sigma$ approximates the unit ball in the sense that 
any point on the surface $\partial P^\sigma$ has radial coordinate $1 + O(s_\sigma^{-1})$.
The polyhedron $P^\sigma$ is moreover symmetric under 
the maps $(k^1,k^2,k^3) \mapsto (\pm k^1, \pm k^2, \pm k^3)$
and ``almost symmetric'' under the maps 
$(k^1,k^2,k^3) \mapsto (k^a, k^b, k^c)$
for $(a,b,c) \ne (1,2,3)$, see \cite[Lemma 2.11]{Lauritsen.Seiringer.2024}.

The \emph{Fermi polyhedron} $P_F^\sigma$ is then defined as 
$P_F^\sigma := k_F^\sigma P^\sigma \cap \frac{2\pi}{L}\Z^3$.

Moreover, $L$ is chosen large such that $\frac{k_F^\sigma L}{2\pi}$ is rational and large for $\sigma \in \{\uparrow,\downarrow\}$.
\end{defn}

\begin{remark}\label{rmk.k_F.rational}
We choose $k_F^\sigma$ such that $k_F^\uparrow / k_F^\downarrow$ is rational since we need $L$ with $\frac{k_F^\sigma L}{2\pi}$ 
rational for both values of $\sigma \in \{\uparrow,\downarrow\}$. 
\end{remark}

\begin{remark}\label{rmk.dependent.parameters}
The free parameters are the \emph{Fermi momenta} $k_F^{\sigma}$, the length of the box $L$ and the number of corners of the polyhedra $s_\sigma$.
The particle numbers are then $N_\sigma = \# P_F^\sigma$ and the particle densities are 
$\rho_\sigma = N_\sigma/L^3 = \frac{1}{6\pi^2} (k_F^\sigma)^{3}\left(1 + O(N_\sigma^{-1/3})\right)$.
Not all densities $\rho_{\sigma 0}$ arise this way. 
We need some argument to consider general densities $\rho_{\sigma 0}$.
This is discussed in \Cref{sec.box}.
Essentially by continuity and density of the rationals in the reals we can extend results for the densities 
arising as $\rho_\sigma = N_\sigma/L^3$ to general densities $\rho_{\sigma 0}$.

We will later choose $L, s_\sigma$ depending on $a^3 \rho$, meaning more precisely on $(k_F^\uparrow + k_F^\downarrow)a$,
such that $L, s_\sigma \to \infty$ as $a^3 \rho \to 0$. 
Concretely we will choose $s_\sigma \sim (a^3\rho)^{-1/3 + \eps}$ for some small $\eps > 0$.
\end{remark}

\noindent
Next, we recall some properties of the Fermi polyhedron from \cite{Lauritsen.Seiringer.2024}.
For the kinetic energy (density) of the Slater determinants we have by
\cite[Lemma 2.13]{Lauritsen.Seiringer.2024}
\begin{equation}\label{eqn.energy.fermi.polyhedron}
    \frac{1}{L^3 }\sum_{k\in P_F^\sigma} |k|^2 = \frac{3}{5} (6\pi^2)^{2/3} \rho_\sigma^{5/3} 
    (1 + O(s_\sigma^{-2}) + O(N_\sigma^{-1/3})).
\end{equation}
Here the $s_\sigma$-dependent error is only negligible if we take $s_\sigma$ large enough --- we need that the Fermi polyhedron
approximates the Fermi ball well in order for the kinetic energies (of the associated Slater determinants) to be close. 
(Recall that the Slater determinant with momenta in the Fermi ball is the ground state of the non-interacting system.)

Moreover, for $N_\sigma = \#P_F^\sigma$ sufficiently large,
the Fermi polyhedron satisfies the following bounds by 
\cite[Lemmas 2.12 and 4.9]{Lauritsen.Seiringer.2024} (see also \cite{Kolomoitsev.Lomako.2018}).
\begin{subequations}\label{eqn.derivative.lebesgue.constants}
\begin{align}
\int_\Lambda \abs{\frac{1}{L^3}\sum_{k\in P_F^\sigma} e^{ikx}} \ud x 
& \leq C s_\sigma (\log N_\sigma)^3 
\leq C s (\log N)^3,
\label{eqn.derivative.lebesgue.constant.0}
\\
\int_\Lambda \abs{\frac{1}{L^3}\sum_{k\in P_F^\sigma} k^j e^{ikx}} \ud x
& \leq C s_\sigma \rho_\sigma^{1/3} (\log N_\sigma)^3
\leq C s \rho^{1/3} (\log N)^3,
\label{eqn.derivative.lebesgue.constant.1}
\\
\int_\Lambda \abs{\frac{1}{L^3}\sum_{k\in P_F^\sigma} k^j k^{j'} e^{ikx}} \ud x
& \leq C s_\sigma \rho_\sigma^{2/3} (\log N_\sigma)^4
\leq C s \rho^{2/3} (\log N)^4,
\label{eqn.derivative.lebesgue.constant.2}
\end{align}
\end{subequations}
for any $j, j' = 1,2,3$ where $s = \max\{s_\uparrow, s_\downarrow\}$, $\rho = \rho_\uparrow + \rho_\downarrow$, $N = N_\uparrow + N_\downarrow$
and $k^j$ denotes the $j$'th component of the vector $k=(k^1,k^2,k^3)$.

The first bound, \Cref{eqn.derivative.lebesgue.constant.0}, is needed to prove the absolute convergence of the Gaudin--Gillespie--Ripka expansion 
discussed in \Cref{sec.gaudin,sec.abs.conv}.
The second two bounds, \Cref{eqn.derivative.lebesgue.constant.1,eqn.derivative.lebesgue.constant.2}, are needed 
to bound the terms with $\rho^{(2,0)}_{\textnormal{Jas}}$ and $\rho^{(0,2)}_{\textnormal{Jas}}$ in \Cref{eqn.energy.compute}. 
More precisely they are used in the proof of \Cref{lem.(2.0).density} below, but only then.


Finally, we recall that the scattering functions satisfy the scattering equations
(Euler-Lagrange equations of the defining minimization problems in \Cref{def.scattering.length})
\begin{equation}\label{eqn.f.scat}
  - 2\Delta f_{s0} + v f_{s0} = 0,
  \qquad 
  -4 x \cdot \nabla f_{p0} - 2|x|^2 \Delta f_{p0} + |x|^2 v f_{p0} = 0.
\end{equation}
Moreover
\begin{lemma}[{\cite[Appendix A]{Lieb.Yngvason.2001}, see also \cite[Lemma 2.2]{Lauritsen.Seiringer.2024}}]
\label{lem.properties.scattering.function}
The functions $f_{s0}$ and $f_{p0}$ are real-valued and radial. Moreover 
\[
\left[ 1 - \frac{a}{|x|}\right]_+ \leq f_{s0}(x) \leq 1,
    \qquad 
\left[ 1 - \frac{a_p^3}{|x|^3}\right]_+ \leq f_{p0}(x) \leq 1.
\]
For $|x|\geq R_0$, the range of $v$, the left-hand-sides are equalities.
\end{lemma}

\section{Gaudin-Gillespie-Ripka expansion}\label{sec.gaudin}
In this section we calculate reduced densities of the trial state $\psi_{N_\uparrow,N_\downarrow}$.
The ideas behind this calculation are mostly contained in (the formal calculations of) \cite{Gaudin.Gillespie.ea.1971}. 
The calculation we give here is a slight generalization thereof including the spin.
Additionally, we give  conditions for the final formulas (given in \Cref{thm.gaudin}) 
to hold, i.e. we give conditions for their absolute convergence. 
The argument here is in spirit the same as that of \cite[Section 3]{Lauritsen.Seiringer.2024}.
Here it is slightly more involved as we have to take into account the different spins.
In \cite[Section 3]{Lauritsen.Seiringer.2024} there is only one value of the spin.

In the calculations below one may replace the functions $f_s, f_p$ and the one-particle density matrices $\gamma_{N_\sigma}^{(1)}$
by more general functions. We discuss this in \Cref{rmk.general.f.gamma} below.

\subsection{Calculation of the normalization constant}\label{sec.calc.C_N}
We first compute the normalization constant $C_{N_\uparrow,N_\downarrow}$.
Recall the definition of the trial state $\psi_{N_\uparrow,N_\downarrow}$ in \Cref{eqn.trial.state}.
Write $f_{\mu\nu}^2 = 1 + g_{\mu\nu}$ for all the $f$-factors 
and factor out the product $\prod_{\mu < \nu} f_{\mu\nu}^2 = \prod_{\mu < \nu} (1 + g_{\mu\nu})$.
We are then led to define the set $\mcG_{p,q}$ as the set of all graphs on $p$ black and $q$ white vertices 
such that each vertex has degree at least $1$, i.e. has an incident edge.
We label the black vertices as $V_p^\uparrow = \{(1,\uparrow),\ldots,(p,\uparrow)\}$ and 
the white vertices as 
$V_q^\downarrow = \{(1,\downarrow),\ldots,(q,\downarrow)\}$. 
For an edge $e=(\mu,\nu)$ we write $g_e = g_{\mu\nu}$
and define 
\[
  W_{p,q} 
  = W_{p,q}(X_p,Y_q)
  = \sum_{G\in \mcG_{p,q}} \prod_{e\in G} g_e.
\]
Then 
\[
\begin{aligned}
  C_{N_\uparrow,{N_\downarrow}}
  & = \idotsint \prod_{\mu<\nu} (1+g_{\mu\nu}) |D_{N_\uparrow}|^2 |D_{N_\downarrow}|^2 \ud X_{N_\uparrow} \ud Y_{N_\downarrow}
  \\
  & = 
  \idotsint 
  \left[
  1 + \sum_{\substack{0\leq p \leq N_\uparrow \\ 0 \leq q \leq {N_\downarrow} \\ p + q \geq 2}}
  \frac{N_\uparrow(N_\uparrow - 1)\cdots (N_\uparrow - p + 1) N_\downarrow (N_\downarrow - 1) \cdots (N_\downarrow - q + 1)}{p!q!}
  W_{p,q} 
  \right]
  \\ & \quad \times 
  |D_{N_\uparrow}|^2 |D_{N_\downarrow}|^2 
  \ud X_{N_\uparrow} \ud Y_{N_\downarrow}
  \\
  & = N_\uparrow! {N_\downarrow}! \left[
    1 + \sum_{\substack{0\leq p \leq N_\uparrow \\ 0 \leq q \leq {N_\downarrow} \\ p + q \geq 2}}
    \frac{1}{p!q!} \idotsint W_{p,q} \rho^{(p,q)} \ud X_p \ud Y_q
    \right].
\end{aligned}
\]
(Recall the definition of $\rho^{(p,q)}$ in \Cref{eqn.normalize.rho.nm}.)
A simple calculation using the Wick rule then shows (recall the definition of $\gamma_{\mu\nu}$ in \Cref{eqn.def.gamma.mu.nu})
\[
  \rho^{(p,q)}(X_p,Y_q)
  = \det \left[\gamma_{\mu\nu}\right]_{\mu,\nu \in V_{p,q}}
  = \det[\gamma_{N_\uparrow}^{(1)}(x_i;y_j)]_{1\leq i,j\leq p}
  \det[\gamma_{N_\downarrow}^{(1)}(y_i;y_j)]_{1\leq i,j\leq q}
\]
Taking this determinantal expression as the definition 
we have $\rho^{(p,q)} = 0$ for $p > N_\uparrow$ or $q > {N_\downarrow}$ since the matrices
$[\gamma_{(i,\uparrow),(j,\uparrow)}]_{i,j \in \N}$ and 
$[\gamma_{(i,\downarrow),(j,\downarrow)}]_{i,j \in \N}$ have ranks $N_\uparrow$ and ${N_\downarrow}$ respectively.
Thus we may extend the $p$- and $q$-sums to $\infty$.
Now, expanding the determinant $\rho^{(p,q)}$ and the $W_{p,q}$ we group the permutations and the graph together in a diagram.
We will for the calculation of the reduced densities need a slightly more general definition, which we now give.

\begin{defn}\label{def.diagrams}
The set $\mcG_{p,q}^{n,m}$ is the set of all graphs with $p$ \emph{internal} black vertices, 
$n$ \emph{external} black vertices, $q$ \emph{internal} white vertices and $m$ \emph{external} white vertices, 
such that there are no edges between \emph{external} vertices, and such that all \emph{internal} vertices has degree at least $1$.
That is, all \emph{internal} vertices are incident to at least one edge and \emph{external} vertices may have degree $0$.
As above we label the black vertices as $V_{p+n}^{\uparrow} = \{(1,\uparrow),\ldots,(p+n,\uparrow)\}$ where the first $n$ are the \emph{external} vertices.
The white vertices are labelled $V_{q+m}^{\downarrow} = \{(1,\downarrow),\ldots,(q+m,\downarrow)\}$, where the first $m$ are the \emph{external} vertices. 
In case $n=m=0$ we recover $\mcG_{p,q}^{0,0} = \mcG_{p,q}$.

If we need the vertices to have certain labels we will write 
$\mcG_{B,W}^{B^*,W^*}$ (or similar with only some of $p,q,n,m$ replaced by sets) 
for the set of all graphs with \emph{internal} black vertices $B$, 
\emph{external} black vertices $B^*$, \emph{internal} white vertices $W$ and \emph{external} white vertices $W^*$,
where $B,B^* \subset V_\infty^\uparrow$ and $W,W^* \subset V_\infty^\downarrow$ are all pairwise disjoint.

The set $\mcD_{p,q}^{n,m}$ is the set of all \emph{diagrams} on $p$ \emph{internal} black vertices, 
$n$ \emph{external} black vertices, $q$ \emph{internal} white vertices and $m$ \emph{external} white vertices.
Such a diagram is a tuple $D = (\pi, \tau, G)$ where $\pi\in \mcS_{p+n}$, $\tau\in \mcS_{q+m}$ 
(viewed as directed graphs on the black and white vertices respectively) 
and $G\in \mcG_{p,q}^{n,m}$.

We will refer to the edges in $G$ as $g$-edges and the graph $G$ as a $g$-graph. 
Moreover, we refer to the edges in both $\pi$ and $\tau$ as $\gamma$-edges.

The value of the diagram $D = (\pi, \tau, G) \in \mcD_{p,q}^{n,m}$ is the function 
\[
\begin{aligned}
  & 
  \Gamma_{D}^{n,m}(X_n, Y_m)
    \\ & \quad 
      = 
      (-1)^{\pi}(-1)^{\tau} \idotsint 
      \prod_{e\in G} g_e \prod_{i=1}^{p+n} \gamma_N^{(1)}(x_i;x_{\pi(i)})
      \prod_{j=1}^{q+m} \gamma_{{N_\downarrow}}^{(1)}(y_j; y_{\tau(j)}) 
      \ud X_{[n+1,n+p]} \ud Y_{[m+1,m+q]}.
\end{aligned}
\]
If $p=0$ and/or $q=0$ there are no integrations in the $x_i$ and/or $y_j$ variables.

A diagram $D = (\pi, \tau, G)$ is said to be \emph{linked} if the graph $\tilde G$ with union all edges of $\pi, \tau$ and $G$ is connected.
The set of linked diagrams is denoted $\mcL_{p,q}^{n,m}$.

In case $m=n=0$ we write $\mcD_{p,q}^{0,0} = \mcD_{p,q}, \mcL_{p,q}^{0,0} = \mcL_{p,q}$ and $\Gamma_D^{0,0} = \Gamma_D$ (i.e. without a superscript).
\end{defn}

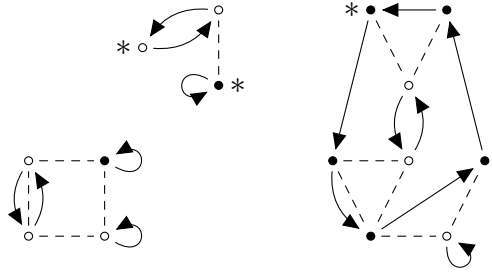
\begin{figure}[htb]
\centering
\begin{tikzpicture}[line cap=round,line join=round,>=triangle 45,x=1.0cm,y=1.0cm]
\node (w1) at (0,1) {};
\node (w3) at (0,2) {};
\node (w4) at (1,1) {};
\node (b2) at (1,2) {};
\draw[dashed] (w1) -- (w3) -- (b2);
\draw[dashed] (w1) -- (w4) -- (b2);
\draw[->] (w1) to[bend right] (w3);
\draw[->] (w3) to[bend right] (w1);
\draw[->] (w4) to[out=-30,in=30,loop] ();
\draw[->] (b2) to[out=-30,in=30,loop] ();
\node (w2) at (1.5,3.5) {};
\node (b1) at (2.5,3) {};
\node (w5) at (2.5,4) {};
\draw[dashed] (b1) -- (w5);
\draw[->] (w2) to[bend right] (w5);
\draw[->] (w5) to[bend right] (w2);
\draw[->] (b1) to[out=150,in=-150,loop] ();
\node (b3) at (4,2) {};
\node (b4) at (4.5,1) {};
\node (w6) at (5,2) {};
\node (w7) at (5.5,1) {};
\node (b5) at (6,2) {};
\draw[dashed] (b3) -- (w6) -- (b4) -- (w7) -- (b5);
\draw[dashed] (b3) -- (b4);
\node (b6) at (4.5,4) {};
\node (b7) at (5.5,4) {};
\node (w8) at (5,3) {};
\draw[dashed] (b6) -- (w8) -- (b7);
\draw[->] (w6) to[bend right] (w8);
\draw[->] (w8) to[bend right] (w6);
\draw[->] (w7) to[out=-90,in=-30,loop] ();
\draw[->] (b5) to (b7);
\draw[->] (b7) to (b6);
\draw[->] (b6) to (b3);
\draw[->] (b3) to[bend right] (b4);
\draw[->] (b4) to (b5);
\foreach \i in {1,...,8} \draw (w\i) circle [radius=1.5pt];
\foreach \i in {1,...,7} \draw[fill] (b\i) circle [radius=1.5pt];
\node[anchor = east] at (w2) {$*$};
\node[anchor = west] at (b1) {$*$};
\node[anchor = east] at (b6) {$*$};
\end{tikzpicture}
\caption{Example of a diagram $(\pi, \tau, G)$ with $3$ linked components. 
Vertices denoted by $\bullet$ are the black vertices, i.e. of spin $\uparrow$, 
and vertices denoted by $\circ$ are the white vertices, i.e. of spin $\downarrow$.
Moreover, vertices with label $*$ are \emph{external},
dashed lines denote $g$-edges and arrows $(\mu,\nu) = ( (i,\sigma), (j,\sigma'))$ 
denote $\gamma$-edges, i.e. that $\pi(i)=j$ if $\sigma=\sigma' = \uparrow$ 
or $\tau(i)=j$ if $\sigma=\sigma' = \downarrow$.
Note that there are no $\gamma$-edges between vertices of different colours (i.e. with different spin).}
\end{figure}

\noindent
In terms of diagrams we have 
\begin{equation}\label{eqn.C_N.diagrams}
  \begin{aligned}
  C_{N_\uparrow,{N_\downarrow}} 
  & = N_\uparrow! {N_\downarrow}! \left[1 + \sum_{\substack{p,q \geq 0 \\ p+q\geq 2}} \frac{1}{p!q!} \sum_{D\in \mcD_{p,q}} \Gamma_D\right].
  \end{aligned}
\end{equation}
We may decompose any diagram $D = (\pi,\tau,G)$  into its linked components. 
For this, note that its value $\Gamma_D$ factors over its linked components. 
Moreover, each linked component has at least $2$ vertices, since they have degree at least one.
Thus, 
\[
  \begin{aligned}
& 
\frac{1}{p!q!} \sum_{D\in \mcD_{p,q}} \Gamma_D
\\ & \quad =
  \underbrace{\sum_{k=1}^\infty}_{\textnormal{\# lnk. cps.}} \frac{1}{k!}
  \underbrace{
  \sum_{\substack{p_1,q_1 \geq 0 \\ p_1+q_1\geq 2}}
  \cdots 
  \sum_{\substack{p_k,q_k \geq 0 \\ p_k+q_k\geq 2}}
  }_{\textnormal{sizes linked components}}
  \chi_{(\sum p_\ell = p)} \chi_{(\sum q_\ell = q)}
  \underbrace{
  \sum_{D_1\in \mcL_{p_1,q_1}} 
  \cdots 
  \sum_{D_k\in \mcL_{p_k,q_k}} 
  }_{\textnormal{linked components}}
  \frac{\Gamma_{D_1}}{p_1!q_1!} \cdots   \frac{\Gamma_{D_k}}{p_k!q_k!}.
  \end{aligned}
\]
Here the factor $\frac{1}{k!}$ comes from counting the possible ways to label the 
$k$ linked components and the factors $\frac{1}{p_\ell!q_\ell!}$ come from counting 
the possible ways of labelling the vertices in the different linked components (and using the factor $\frac{1}{p!q!}$ already present). 
If we assume that the sum $\sum_{p,q: p+q\geq 2}\frac{1}{p!q!}\sum_{D\in \mcL_{p,q}}\Gamma_{D}$
is absolutely convergent, (more precisely we assume that 
$\sum_{p,q: p+q\geq 2}\frac{1}{p!q!}\abs{\sum_{D\in \mcL_{p,q}}\Gamma_{D}} < \infty$,)
then we may interchange the $p,q$-sum with the $p_\ell,q_\ell$-sums.
The absolute convergence is proven in \Cref{thm.gaudin} below.
Thus, under the conditions of \Cref{thm.gaudin}, we have
\begin{equation}\label{eqn.formula.C_N.linked}
  \begin{aligned}
  C_{N_\uparrow,{N_\downarrow}} 
  & = N_\uparrow!{N_\downarrow}!
  \left[
  1 + 
  \sum_{k=1}^\infty\frac{1}{k!}
  \sum_{\substack{p_1,q_1 \geq 0\\ p_1+q_1\geq 2}} \cdots \sum_{\substack{p_k,q_k \geq 0\\ p_k+q_k\geq 2}}
  \sum_{D_1\in \mcL_{p_1,q_1}} 
  \cdots 
  \sum_{D_k\in \mcL_{p_k,q_k}} 
  \frac{\Gamma_{D_1}}{p_1!q_1!} \cdots   \frac{\Gamma_{D_k}}{p_k!q_k!}
  \right]
  \\ & = 
  N_\uparrow!{N_\downarrow}!\left[1 + 
  \sum_{k=1}^\infty\frac{1}{k!}
  \left(
  \sum_{\substack{p,q \geq 0\\ p+q\geq 2}}
  \frac{1}{p!q!}
  \sum_{D\in \mcL_{p,q}} 
  \Gamma_{D}\right)^k\right]
  \\
  & = N_\uparrow! {N_\downarrow}! \exp \left[\sum_{\substack{p,q \geq 0\\ p+q \geq 2}} \frac{1}{p!q!} \sum_{D\in \mcL_{p,q}} \Gamma_D \right].
  \end{aligned}
\end{equation}

\subsection{Calculation of the reduced densities}\label{sec.calc.rho.nm}
For the calculation of the reduced densities we need to keep track of also the external vertices.
First, we have the formula (for $n+m \geq 1$)
\begin{equation}\label{eqn.rho.nm.calc.initial}
\begin{aligned}
  \rho^{(n,m)}_{\textnormal{Jas}}
  & = N_\uparrow(N_\uparrow-1)\cdots (N_\uparrow-n+1){N_\downarrow}({N_\downarrow}-1)\cdots ({N_\downarrow}-m+1)
  \\ & \quad \times 
    \idotsint |\psi_{N_\uparrow,{N_\downarrow}}(X_{N_\uparrow},Y_{N_\downarrow})|^2 \ud X_{[n+1,N_\uparrow]} \ud Y_{[m+1,{N_\downarrow}]}
  \\ 
  & = \frac{N_\uparrow(N_\uparrow-1)\cdots (N_\uparrow-n+1){N_\downarrow}({N_\downarrow}-1)\cdots ({N_\downarrow}-m+1)}{C_{N_\uparrow,{N_\downarrow}}} 
  \prod_{\substack{\mu < \nu \\ \mu,\nu \in V_{n,m}}} f_{\mu\nu}^2
  \\ & \quad \times 
    \idotsint  
      \prod_{\mu\in V_{n,m}, \nu \notin V_{n,m}} (1 + g_{\mu\nu}) 
      \prod_{\substack{\mu < \nu \\ \mu,\nu \notin V_{n,m}}} (1 + g_{\mu\nu})
      D_{N_\uparrow}(X_{N_\uparrow})D_{N_\downarrow}(Y_{N_\downarrow})
      \ud X_{[n+1,N_\uparrow]} \ud Y_{[m+1,{N_\downarrow}]}
  \\
  & = 
  \frac{N_\uparrow!{N_\downarrow}!}{C_{N_\uparrow,{N_\downarrow}}}
  \prod_{\substack{\mu < \nu \\ \mu,\nu \in V_{n,m}}} f_{\mu\nu}^2
  \left[
  \sum_{\substack{p,q \geq 0}}
  \frac{1}{p!q!} 
    \idotsint 
    \rho^{(n+p,m+q)}
    \sum_{G\in \mcG_{p,q}^{n,m}} 
    \prod_{e\in G} g_e 
     \ud X_{[n+1,n+p]} \ud Y_{[m+1,m+q]}
     \right]
  \\
  & = 
  \frac{N_\uparrow!{N_\downarrow}!}{C_{N_\uparrow,{N_\downarrow}}}
  \prod_{\substack{\mu < \nu \\ \mu,\nu \in V_{n,m}}} f_{\mu\nu}^2
  \left[\rho^{(n,m)} 
  + \sum_{\substack{p,q \geq 0\\ p+q \geq 1}} \frac{1}{p!q!} 
  \sum_{D\in \mcD_{p,q}^{n,m}} \Gamma_D^{n,m}
  \right]
\end{aligned}
\end{equation}
where we extended the $p,q$-sums to $\infty$ as in \Cref{sec.calc.C_N} above
and used that the $p=q=0$ term gives 
\[
  \sum_{D\in \mcD_{0,0}^{n,m}} \Gamma_D^{n,m} = \rho^{(n,m)}.
\]
Note here that the $p,q$-sum does not require $p+q\geq 2$, since the internal vertices may connect to external ones.
As for the normalization constant in \Cref{sec.calc.C_N} we decompose each diagram $D$ into its linked components. 
Here we have to keep track of which linked components contain which external vertices. 
To do this we introduce the set 
\begin{equation}\label{eqn.define.Pi.kappa}
\Pi^{n,m}_\kappa : = \left\{
(\mcB^*, \mcW^*) : 
\begin{matrix}
\mcB^* = (B_1^*,\ldots,B_\kappa^*) \textnormal{ partition of } \{1,\ldots,n\}, 
\\ 
\mcW^* = (W_1^*,\ldots,W_\kappa^*) \textnormal{ partition of } \{1,\ldots,m\},
\\
\textnormal{ For all $\lambda:$ $B_\lambda^* \ne \varnothing$ and/or $W_\lambda^* \ne \varnothing$.}
\end{matrix}
\right\}.
\end{equation}
The set $\Pi_\kappa^{n,m}$ 
parametrizes all possible ways for the diagram $D \in \mcD_{p,q}^{n,m}$ to have exactly $\kappa$ many linked components
containing at least $1$ external vertex each. 
Note that for $\kappa > n+m$ we have $\Pi_\kappa^{n,m} = \varnothing$, since we require that 
for all $\lambda$ we have $B_\lambda^* \ne \varnothing$ or $W_\lambda^* \ne \varnothing$.
Denoting then $k$ the number of linked components with only internal vertices we get the following.
\begin{equation}\label{eqn.decompose.linked.components.rho.nm.initial}
\begin{aligned}
&
\frac{1}{p!q!} 
  \sum_{D\in \mcD_{p,q}^{n,m}} \Gamma_D^{n,m}
\\ & 
= \sum_{k=0}^\infty \frac{1}{k!} 
  \sum_{\kappa=1}^{n+m} \frac{1}{\kappa!}
  \sum_{(\mcB^*, \mcW^*)\in \Pi_\kappa^{n,m}} 
  \sum_{\substack{p_1^*,q_1^* \geq 0}}
  \cdots 
  \sum_{\substack{p_\kappa^*,q_\kappa^* \geq 0}}
  \sum_{\substack{p_1,q_1 \geq 0\\ p_1+q_1\geq 2}}
  \cdots 
  \sum_{\substack{p_k,q_k \geq 0 \\ p_k+q_k\geq 2}}
  \chi_{(\sum_{\lambda} p_\lambda^* + \sum_{\ell} p_\ell = p)}
  \chi_{(\sum_{\lambda} q_\lambda^* + \sum_{\ell} q_\ell = q)}
  \\ & \quad \times 
  \sum_{D_1^* \in \mcL_{p_1^*,q_1^*}^{\#B_1^*,\#W_1^*}}
  \cdots 
  \sum_{D_\kappa^* \in \mcL_{p_\kappa^*,q_\kappa^*}^{\#B_\kappa^*,\# W_\kappa^*}} 
  \frac{\Gamma_{D_1^*}^{\# B_1^*,\#W_1^*}(X_{B_1^*},Y_{W_1^*})}{p_1^*! q_1^*!} 
  \cdots 
  \frac{\Gamma_{D_\kappa^*}^{\# B_\kappa^* ,\# W_\kappa^*}(X_{B_\kappa^*},Y_{W_\kappa^*})}{p_\kappa^*! q_\kappa^*!} 
  \\ & \quad \times 
  \sum_{D_1 \in \mcL_{p_1,q_1}}
  \cdots 
  \sum_{D_k \in \mcL_{p_k,q_k}} 
  \frac{\Gamma_{D_1}}{p_1! q_1!} 
  \cdots 
  \frac{\Gamma_{D_k}}{p_k! q_k!}.  
\end{aligned}
\end{equation}
(Note that the linked components with external vertices may have $0$ or $1$ internal vertices, 
i.e. the $p^*_\lambda, q^*_\lambda$-sums do not require $p^*_\lambda + q^*_\lambda \geq 2$.)
The factorial factors come from counting the different labellings:
The factors $\frac{1}{k!}$ and $\frac{1}{\kappa!}$ from the labellings of the clusters 
and the factors $\frac{1}{p_\lambda^*!}, \frac{1}{q_\lambda^*!},\frac{1}{p_\ell!},\frac{1}{q_\ell!}$
from labelling the internal vertices of the different clusters exactly as in \Cref{sec.calc.C_N} above.

If we assume absolute convergence of all the $\Gamma^{n',m'}$-sums with $n'\leq n$ and $m'\leq m$
(i.e. that 
$ \sum_{p,q\geq0} \frac{1}{p!q!} \abs{\sum_{D\in \mcL_{p,q}^{n',m'}} \Gamma_D^{n',m'}} < \infty $)
then we may interchange the $p,q$-sum with the $p_\lambda^*,q_\lambda^*$- and $p_\ell,q_\ell$-sums.
We then get 
\begin{equation}\label{eqn.calc.Gamma.nm.final}
\begin{aligned}
  & 
  \sum_{\substack{p,q \geq 0}} \frac{1}{p!q!} 
  \sum_{D\in \mcD_{p,q}^{n,m}} \Gamma_D^{n,m}
  \\ & \quad = 
  \sum_{k=0}^\infty \frac{1}{k!} 
  \left(
  \sum_{\substack{p,q \geq 0 \\ p+q\geq 2}}
  \frac{1}{p!q!}
  \sum_{D\in \mcL_{p,q}} 
  \Gamma_{D}\right)^k
  \\ & \qquad \times 
  \sum_{\kappa=1}^{n+m} \frac{1}{\kappa!}
  \sum_{(\mcB^*, \mcW^*)\in \Pi_\kappa^{n,m}} 
  \prod_{\lambda=1}^\kappa
  \left[\sum_{p_\lambda,q_\lambda \geq 0}
      \frac{1}{p_\lambda! q_\lambda!}
      \sum_{D_\lambda \in \mcL_{p_\lambda, q_\lambda}^{\# B_\lambda^*, \# W_\lambda^*}}
      \Gamma_{D_\lambda}^{\# B_\lambda^*,\#W_\lambda^*}(X_{B_\lambda^*}, Y_{W_\lambda^*})\right].
\end{aligned}
\end{equation}
Thus by \Cref{eqn.formula.C_N.linked,eqn.rho.nm.calc.initial} we conclude the formula 
\[
\begin{aligned}
& \rho^{(n,m)}_{\textnormal{Jas}}(X_n,Y_m)
\\ & \quad
  = 
   \left[\prod_{\substack{\mu,\nu \in V_{n,m} \\ \mu < \nu}} f_{\mu\nu}^2 \right]
    \sum_{\kappa=1}^{n+m} \frac{1}{\kappa!} 
    \sum_{(\mcB^*,\mcW^*)\in \Pi_\kappa^{n,m}}
    \prod_{\lambda=1}^\kappa
    \left[\sum_{p_\lambda,q_\lambda \geq 0}
            \frac{1}{p_\lambda! q_\lambda!}
            \sum_{D_\lambda \in \mcL_{p_\lambda, q_\lambda}^{\# B_\lambda^*, \# W_\lambda^*}}
            \Gamma_{D_\lambda}^{\# B_\lambda^*,\# W_\lambda^*}(X_{B_\lambda^*}, Y_{W_\lambda^*})\right]
\end{aligned}
\]
under the assumption of absolute convergence.

\subsection{Summary of results}
With the calculation above we may then state the following theorem.

\begin{thm}\label{thm.gaudin}
For integers $n_0,m_0\geq 0$ there exist constants $c_{n_0,m_0}, C_{n_0,m_0} > 0$ (small and large respectively) such that if 
$s a b^2 \rho (\log N)^3 < c_{n_0,m_0}$ then 
\begin{equation}\label{eqn.thm.gaudin.abs.conv}
  \sum_{\substack{p,q\geq 0 \\ p+q \geq 2}} \frac{1}{p!q!} 
  \abs{\sum_{D \in \mcL_{p, q}}
    \Gamma_D}
  < \infty,
  \qquad
  \sum_{p,q\geq 0} \frac{1}{p!q!} 
  \abs{\sum_{D \in \mcL_{p, q}^{n, m}}
    \Gamma_D^{n,m}}
  \leq C_{n_0,m_0} \rho^{n+m}
  < \infty 
\end{equation}
for any $n\leq n_0$ and $m\leq m_0$ with $n+m\geq 1$.
In particular, then 
\begin{equation}\label{eqn.thm.gaudin.main}
\begin{aligned}
  & \rho^{(n,m)}_{\textnormal{Jas}}(X_n,Y_m)
  \\ & \quad = 
   \left[\prod_{\substack{\mu,\nu \in V_{n,m} \\ \mu < \nu}} f_{\mu\nu}^2\right]
    \sum_{\kappa=1}^{n+m} \frac{1}{\kappa!} 
    \sum_{(\mcB^*,\mcW^*)\in \Pi_\kappa^{n,m}}
    \prod_{\lambda=1}^\kappa
    \left[\sum_{p_\lambda,q_\lambda \geq 0}
            \frac{1}{p_\lambda! q_\lambda!}
            \sum_{D_\lambda \in \mcL_{p_\lambda, q_\lambda}^{\# B_\lambda^*, \# W_\lambda^*}}
            \Gamma_{D_\lambda}^{\# B_\lambda^*,\# W_\lambda^*}(X_{B_\lambda^*}, Y_{W_\lambda^*})\right],
\end{aligned}
\end{equation}
where $\Pi_\kappa^{n,m}$ is defined in \Cref{eqn.define.Pi.kappa}.
\end{thm}

\noindent
As particular cases we note that for $n+m=1$ we have by translation invariance that 
\begin{equation}\label{eqn.rho1.translation.invariance}
    \rho_\uparrow = \rho^{(1,0)}_{\textnormal{Jas}}
    = \sum_{p,q\geq 0}\frac{1}{p!q!} \sum_{D\in \mcL_{p,q}^{1,0}} \Gamma_D^{1,0},
    \qquad 
    \rho_\downarrow = \rho^{(0,1)}_{\textnormal{Jas}}
    = \sum_{p,q\geq 0}\frac{1}{p!q!} \sum_{D\in \mcL_{p,q}^{0,1}} \Gamma_D^{0,1}.
\end{equation}
We give the proof of \Cref{thm.gaudin} below.

\begin{remark}[{Higher spin}]\label{rmk.higher.spin.diagrams.def}
One may readily generalize the computation above to a general number of spins $S$. 
For this one introduces vertices of more colours and diagrams with such, i.e. the sets of graphs and diagrams
$\mcG_{p_1,\ldots,p_S}^{n_1,\ldots,n_S}, \mcD_{p_1,\ldots,p_S}^{n_1,\ldots,n_S}, \mcL_{p_1,\ldots,p_S}^{n_1,\ldots,n_S}$
and the values $\Gamma_{D}^{n_1,\ldots,n_S}$.
The condition of absolute convergence is completely analogous. 
\end{remark}

\begin{remark}\label{rmk.N.dependence.abs.conv}
The condition for the absolute convergence is not uniform in the volume, hence the need for a box method as given in \Cref{sec.box}.
The condition of absolute convergence is additionally the reason for introducing the Fermi polyhedron.
This is discussed in \cite[Remark 3.5]{Lauritsen.Seiringer.2024}.
If one did not introduce the Fermi polyhedron and instead used the Fermi ball 
the factor $s(\log N)^3$ in the assumption of \Cref{thm.gaudin} should be replaced by $N^{1/3}$.
\end{remark}

\begin{remark}[{General $f$ and $\gamma$}]\label{rmk.general.f.gamma}
In the computation above we may replace the specific functions $f_s, f_p$ by more general functions $f_{\sigma \sigma'} = f_{\sigma'\sigma}\geq 0$.
(One then introduces $g_e = f^2_{\sigma\sigma'}(z_i - z_j) - 1$ for $e = ( (i,\sigma), (j,\sigma'))$.)

Moreover, for the absolute convergence we may additionally replace the one-particle densities $\gamma_{N_\sigma}^{(1)}$
by general functions $\gamma_\sigma(z_i - z_j)$.
(One then defines $\gamma_{\mu\nu}$ as in \Cref{eqn.def.gamma.mu.nu} above.)
In the computation above we crucially used that $[\det \gamma_{\mu\nu}]_{\mu,\nu \in V_{p,q}}=0$ for appropriately large $p,q$
in order to extend the $p,q$-sums to $\infty$. If for the general $\gamma_\sigma$ this is not valid, 
this step of the computation above is not valid. 
The rest of the calculation starting from what one gets out of this step is however still valid.
That is, the calculation in \Cref{sec.calc.C_N} is valid starting from \Cref{eqn.C_N.diagrams}
and the calculations in \Cref{eqn.decompose.linked.components.rho.nm.initial,eqn.calc.Gamma.nm.final} in \Cref{sec.calc.rho.nm} are valid.

The statement of the absolute convergence in this more general setting reads 
\begin{lemma}\label{lemma.gaudin.general}
Suppose there exists a constant $C_{\textnormal{TG}} \geq 1$ such that 
\begin{equation}\label{eqn.tree.graph.condition}
  \sup_{\sigma, \sigma'} \sup_{z_1,\ldots,z_q} \prod_{1 \leq i < j \leq q}  f_{\sigma \sigma'}(z_i - z_j)^2
  \leq (C_{\textnormal{TG}})^{q}
  \qquad \textnormal{for any } q\in \N.
\end{equation}
Then 
for integers $n_{1,0},\ldots,n_{S,0}$ 
there exists constants $c_{n_{1,0},\ldots,n_{S,0}}, C_{n_{1,0},\ldots,n_{S,0}} > 0$ such that if
\begin{equation}\label{eqn.general.abs.conv.condition}
  \sup_{\sigma} \sum_{k\in \frac{2\pi}{L}\Z^3} \abs{\hat \gamma_\sigma(k)}
  \times \sup_{\sigma,\sigma'} \int_\Lambda  \abs{f_{\sigma\sigma'}^2 - 1} 
  \times 
  \left[1 + \sup_\sigma \int_\Lambda |\gamma_\sigma| \right]
  < c_{n_{1,0},\ldots,n_{S,0}},
\end{equation}
where $\hat \gamma_\sigma(k) = \frac{1}{L^3}\int_\Lambda \gamma_\sigma(x) e^{-ikx} \ud x$ denotes the Fourier transform,
then 
\begin{equation}\label{eqn.abs.conv.general.f.gamma}
\begin{aligned}
  \sum_{\substack{p_1,\ldots,p_S\geq 0 \\ \sum_\sigma p_\sigma \geq 2}}
  \frac{1}{p_1! \cdots p_S!} \abs{ \sum_{D\in \mcL_{p_1,\ldots,p_S}} \Gamma_D}
  & < \infty,
  \\
  \sum_{\substack{p_1,\ldots,p_S \geq 0}} \frac{1}{p_1! \cdots p_S!} \abs{ \sum_{D\in \mcL_{p_1,\ldots,p_S}^{n_1,\ldots,n_S}} \Gamma_D^{n_1,\ldots,n_S}}
  & \leq C_{n_{1,0},\ldots,n_{S,0}} 
  \left[\sup_{\sigma} \sum_{k\in \frac{2\pi}{L}\Z^3} \abs{\hat \gamma_\sigma(k)}\right]^{\sum_\sigma n_\sigma}
  < \infty
\end{aligned}
\end{equation}
for all $n_\sigma \leq n_{\sigma, 0}$ with $\sum_\sigma n_\sigma \geq 1$.
In particular then
\[
  \mcZ :=
  1 + \sum_{\substack{p_1,\ldots,p_S\geq 0 \\ \sum_\sigma p_\sigma \geq 2}}
  \frac{1}{p_1! \cdots p_S!} \sum_{D\in \mcD_{p_1,\ldots,p_S}} \Gamma_D
  = 
  \exp \left[
  \sum_{\substack{p_1,\ldots,p_S\geq 0 \\ \sum_\sigma p_\sigma \geq 2}}
  \frac{1}{p_1! \cdots p_S!} \sum_{D\in \mcL_{p_1,\ldots,p_S}} \Gamma_D
  \right]
\]
and
\begin{multline*} 
  \frac{1}{\mcZ}\sum_{\substack{p_1,\ldots,p_S\geq 0}} 
  \frac{1}{\prod_\sigma p_\sigma!} \sum_{D\in \mcD_{p_1,\ldots,p_S}^{n_1,\ldots,n_S}} 
  \Gamma_D^{n_1,\ldots,n_S}( (X^\sigma_{n_\sigma})_{\sigma=1,\ldots,S})
  \\ 
  = 
  \sum_{\kappa=1}^{\sum_\sigma n_\sigma} \frac{1}{\kappa!}
  \sum_{(\mcV^{*1},\ldots,\mcV^{*S})\in \Pi_\kappa^{n_1,\ldots,n_S}} 
  \prod_{\lambda=1}^\kappa
  \left[\sum_{p_\lambda^1,\ldots,p_\lambda^S \geq 0}
      \frac{1}{\prod_\sigma p_\lambda^\sigma!}
      \sum_{D_\lambda \in \mcL_{p_\lambda^1,\ldots,p_\lambda^S}^{\# V_\lambda^{*1}, \ldots, \# V_\lambda^{*S}}}
      \Gamma_{D_\lambda}^{\# V_\lambda^{*1}, \ldots, \# V_\lambda^{*S}}( (X_{V^{*\sigma}_\lambda}^\sigma)_{\sigma = 1,\ldots,S})
      \right],
\end{multline*}
where
\begin{equation*}
\Pi^{n_1,\ldots,n_S}_\kappa : = 
\left\{
(\mcV^{*1},\ldots,\mcV^{*S}) : 
\begin{matrix}
\mcV^{*\sigma} = (V_1^{*\sigma},\ldots,V_\kappa^{*\sigma}) \textnormal{ partition of } \{1,\ldots,n_\sigma\}
\\
\textnormal{ For all $\lambda:$ $V_\lambda^{*\sigma} \ne \varnothing$ for some $\sigma$}
\end{matrix}
\right\}
\end{equation*}
parametrizes the ways for the external vertices to lie in $\kappa$ different linked components,
the coordinates of each spin $\sigma$ are labelled $x^\sigma_j,  j\in \N$, 
and we denote by $X^\sigma_A = (x^\sigma_j)_{j\in A}$
the coordinates with labels in the set $A$.
\end{lemma}

\noindent
The condition in \Cref{eqn.tree.graph.condition} is the ``stability condition'' of the tree-graph bound 
\cites[Proposition 6.1]{Poghosyan.Ueltschi.2009}{Ueltschi.2018}.
\end{remark}

\noindent
We give the proof of \Cref{lemma.gaudin.general} in \Cref{sec.abs.conv} below for the case $S=2$.
The proof for general $S$ is a straightforward modification, but notationally more cumbersome.
The case $S=1$ is treated in \cite[Section 3.1]{Lauritsen.Seiringer.2024}.
\Cref{thm.gaudin} is an immediate corollary.

\begin{proof}[{Proof of \Cref{thm.gaudin}}]
Note that $f_s, f_p\leq 1$ and 
$\hat\gamma_{N_\sigma}(k) := L^{-3} \int_\Lambda \gamma_{N_\sigma}^{(1)}(x;0) e^{-ikx} \ud x = L^{-3} \chi_{(k\in P_F^\sigma)}$
so $\sum_{k\in \frac{2\pi}{L}\Z^3} \abs{\hat\gamma_{N_\sigma}(k)} = \rho_\sigma \leq \rho$.
Moreover, we have the bounds 
\begin{equation}\label{eqn.bound.int.g}
  \int |g_s| \leq C a b^2, 
  \qquad \int |g_p| \leq C a_p^3 \log(b/a_p) \leq C a b^2,
\end{equation}
which follow by a simple computation using \Cref{lem.properties.scattering.function}.
Recalling also \Cref{eqn.derivative.lebesgue.constants} then \Cref{lemma.gaudin.general} proves the desired. 
\end{proof}

\section{Absolute convergence of the Gaudin-Gillespie-Ripka expansion}\label{sec.abs.conv}
In this section we give the proof of \Cref{lemma.gaudin.general} for the case $S=2$.
The proof is similar to that of \cite[Theorem 3.4]{Lauritsen.Seiringer.2024}.
We need to prove 
(for all $n,m$ and uniformly in $X_n,Y_m$)
\Cref{eqn.abs.conv.general.f.gamma} if \Cref{eqn.tree.graph.condition,eqn.general.abs.conv.condition}
are satisfied.
To simplify notation we define 
\[
  \gamma_\infty := \sup_{\sigma} \sum_{k\in \frac{2\pi}{L}\Z^3} \abs{\hat \gamma_\sigma(k)},
  \qquad 
  I_g := \sup_{\sigma, \sigma'} \int_\Lambda \abs{f_{\sigma \sigma'}^2 - 1} = \sup_e \int_\Lambda |g_e|,
  \qquad
  I_\gamma:=  \sup_{\sigma} \int_\Lambda \abs{\gamma_\sigma},
\]
where as above $\hat \gamma_\sigma(k) = L^{-3} \int_\Lambda \gamma_\sigma(x) e^{-ikx} \ud x$.
\Cref{eqn.general.abs.conv.condition} then reads that $\gamma_\infty I_g (1+I_\gamma)$ is sufficiently small.

We give the proof in two steps. 
First we consider the case $n=m=0$.

\subsection{Absolute convergence of the \texorpdfstring{$\Gamma$}{Gamma}-sum}\label{sec.Gamma.abs.conv}
In this section we show that 
\[
  \sum_{\substack{p,q\geq 0 \\ p+q \geq 2}} \frac{1}{p!q!} 
  \abs{\sum_{D \in \mcL_{p, q}}
    \Gamma_D}
  < \infty
\]
under the relevant conditions.
Defining clusters as connected components of $G$ we split the sum into clusters as in \cite[Section 3.1]{Lauritsen.Seiringer.2024}.
Denoting the sizes of the clusters by $(n_\ell, m_\ell)$, $\ell=1,\ldots,k$
(meaning that the cluster $\ell$ has $n_\ell$ black vertices and $m_\ell$ white vertices)
we get 
\begin{equation}\label{eqn.decompose.clusters}
\begin{aligned}
& 
  \frac{1}{p!q!} 
  \sum_{D \in \mcL_{p, q}}
    \Gamma_D
  \\ &  =   
    \sum_{k=1}^\infty \frac{1}{k!} 
    \sum_{\substack{n_1,\ldots,n_k\geq 0 \\ m_1,\ldots,m_k\geq 0 \\ \textnormal{For each $\ell$: } n_\ell + m_\ell \geq 2}}
    \chi_{(\sum_{\ell} n_\ell = p)}
    \chi_{(\sum_{\ell} m_\ell = q)}
    \frac{1}{\prod_{\ell=1}^k n_\ell! m_\ell!}
    \sum_{G_\ell \in \mcC_{n_\ell,m_\ell}}
    \idotsint 
    \ud X_{p} \ud Y_{q}
    \left[
    \prod_{\ell=1}^k \prod_{e\in G_\ell} g_e
    \right]
  \\ & \quad \times 
    \left[
    \sum_{\substack{\pi \in \mcS_{p} \\ \tau \in \mcS_{q}}}
    (-1)^\pi (-1)^\tau 
    \chi_{( (\pi, \tau, \cup_\ell G_\ell) \textnormal{ linked})}
    \prod_{i=1}^{p} \gamma_{\uparrow}(x_i - x_{\pi(i)})
    \prod_{j=1}^{q} \gamma_{\downarrow}(y_j - y_{\tau(j)})
    \right],
\end{aligned}
\end{equation}
where $\mcC_{p,q}\subset \mcG_{p,q}$ denotes the subset of connected graphs.
The factorial factors arise from counting the possible labellings exactly as in \Cref{sec.gaudin}.

The last line of \Cref{eqn.decompose.clusters} 
is what we will call the \emph{truncated correlation}.
We give a slightly more general definition for later use.
\begin{defn}\label{def.truncated.correlations}
    Let $B_1,\ldots, B_k$ and $W_1,\ldots, W_k$ be sets of distinct 
    black and white vertices respectively, 
    such that for each $\ell=1,\ldots,k$ we have 
    $B_\ell \ne \varnothing$ and/or $W_\ell\ne \varnothing$.
    Then the \emph{truncated correlation}\footnote{The truncated correlation is also sometimes referred to as the \emph{connected correlation}. 
    In particular, this is the terminology used in \cite{Giuliani.Mastropietro.ea.2021}.} is defined as follows.
\begin{equation}\label{eqn.define.rhot}
    \rho_t^{((B_1,W_1),\ldots,(B_k,W_k))}
    = 
    \sum_{\substack{\pi \in \mcS_{\cup_\ell B_\ell} \\ \tau \in \mcS_{\cup_\ell  W_\ell}}}
    (-1)^\pi (-1)^\tau 
    \chi_{( (\pi, \tau, \cup_\ell G_\ell) \textnormal{ linked})}
    \prod_{i\in \cup_\ell B_\ell}  \gamma_{\uparrow}(x_i - x_{\pi(i)})
    \prod_{j\in\cup_\ell W_\ell} \gamma_{\downarrow}(y_j - y_{\tau(j)})
\end{equation}
for any connected graphs $G_\ell \in \mcC_{B_\ell, W_\ell}$.
The definition does not depend on the choice of the graphs $G_\ell$.

If the underlying sets $B_1,\ldots, B_k, W_1, \ldots, W_k$ are clear we will 
also use the notation
\[
    \rho_t^{((\# B_1 , \# W_1), \ldots, (\# B_k, \# W_k))}
    = 
    \rho_t^{((B_1,W_1),\ldots,(B_k,W_k))}.
\]
\end{defn}

\noindent
The truncated correlations are studied in \cite[Appendix D]{Giuliani.Mastropietro.ea.2021}.
To better compare to the definition in \cite{Giuliani.Mastropietro.ea.2021} we note the following.

In \Cref{eqn.define.rhot} we may view $(\pi,\tau)$ together as a permutation of all the vertices (both black and white). 
Moreover, if we instead sum over all permutations $\pi'\in \mcS_{\cup_\ell B_\ell \cup \cup_\ell W_\ell}$
we have that any $\pi'$ not coming from two permutations $\pi, \tau$ on the black (respectively white) vertices contributes $0$,
since any $\gamma$-factor between vertices of different spins is $0$.
That is,
\[
    \rho_t^{((B_1,W_1),\ldots,(B_k,W_k))}
    = \sum_{\pi' \in \mcS_{\cup_\ell B_\ell \cup \cup_\ell W_\ell}}
    (-1)^\pi \chi_{(\pi', \cup G_\ell) \textnormal{ linked})}
    \prod_{\mu \in \cup B_\ell \cup \cup W_\ell} \gamma_{\mu, \pi'(\mu)}.
\]
In \cite[Equation (D.53)]{Giuliani.Mastropietro.ea.2021} is shown the formula for the truncated correlation
\begin{equation}\label{eqn.BBF.rho_t}
    \rho_t^{((B_1,W_1),\ldots,(B_k,W_k))}
    = \sum_{A \in \mcA^{((B_1,W_1),\ldots,(B_k,W_k))}}
    \prod_{(\mu,\nu)\in A} \gamma_{\mu\nu} 
    \int \ud\mu_A(r) \det \mcR(r),
\end{equation}
where $\mcA$ denotes the set of anchored trees, $\mu_A$ is a probability measure and $\mcR(r)$ is an explicit matrix.
The set $\mcA^{((B_1,W_1),\ldots,(B_k,W_k))}$
of anchored trees is the set of all directed graphs on the vertices 
$\cup_\ell B_\ell \cup\cup_\ell W_\ell$ such that each vertex has at most one incoming and at most one outgoing edge, 
and such that upon identifying all vertices inside each cluster
the resulting graph is a (directed) tree.
The matrix $\mcR(r)$ satisfies the bound 
\begin{equation}\label{eqn.bound.det.N(r)}
    \abs{\det \mcR(r)}
    \leq \gamma_\infty^{\sum_\ell (\# B_\ell + \# W_\ell) - (k - 1)}.
\end{equation}
This follows from \cite[Equation (D.57)]{Giuliani.Mastropietro.ea.2021}. 
We give a sketch of the argument here, see also \cite[Lemma D.2]{Giuliani.Mastropietro.ea.2021} and \cite[Lemma 3.10]{Lauritsen.Seiringer.2024}.
\begin{proof}[{Proof (sketch) of \Cref{eqn.bound.det.N(r)}}]
Write $\gamma_\sigma(z_\mu - z_\nu) = \ip{\alpha_\mu}{\beta_\nu}_{\ell^2(\frac{2\pi}{L}\Z^3)}$,
where for $k\in \frac{2\pi}{L}\Z^3$
\[
  \alpha_\mu(k) = e^{-ikz_\mu} \abs{\hat \gamma_\sigma(k)}^{1/2} \frac{\hat \gamma_\sigma(k)}{\abs{\hat \gamma_\sigma(k)}},
  \qquad 
  \beta_\nu(k) = e^{-ikz_\nu} \abs{\hat \gamma_\sigma(k)}^{1/2},
\]
with $\hat \gamma_\sigma(k) = L^{-3} \int_\Lambda \gamma_\sigma(x) e^{-ikx}\ud x$ the Fourier coefficients.
Then by the Gram-Hadamard inequality \cite[Lemma D.1]{Giuliani.Mastropietro.ea.2021} 
\[
\abs{\det [\gamma_\sigma(z_\mu - z_\nu)]_{\mu,\nu \in V_p^\sigma}} 
\leq \prod_{\mu \in V_p^\sigma} \norm{\alpha_\mu}_{\ell^2(\frac{2\pi}{L}\Z^3)} \norm{\beta_\mu}_{\ell^2(\frac{2\pi}{L}\Z^3)}
\leq \left[\sum_{k\in \frac{2\pi}{L}\Z^3} \abs{\hat \gamma(k)}\right]^p.
\]
It is then explained in the proof of \cite[Lemma D.6]{Giuliani.Mastropietro.ea.2021}
how to adapt this argument to bound $\det \mcR(r)$.
\end{proof}

\noindent
Combining \Cref{eqn.bound.det.N(r),eqn.BBF.rho_t} we conclude the bound  
 \begin{equation}\label{eqn.bound.truncated.correlation}
 \abs{\rho_t^{((B_1,W_1),\ldots,(B_k,W_k))}}
 \leq \gamma_\infty^{\sum_\ell (\# B_\ell  + \# W_\ell) - (k - 1)}
 \sum_{A \in \mcA^{((B_1,W_1),\ldots,(B_k,W_k))}}
    \prod_{(\mu,\nu)\in A} \abs{\gamma_{\mu\nu}}.
 \end{equation}

\noindent
With the truncated correlation we may write the last line of \Cref{eqn.decompose.clusters} as 
$\rho_t^{(\mcN, \mcM)}$, where 
\[
  \mcN = (n_1,\ldots,n_k),
  \qquad 
  \mcM = (m_1,\ldots,m_k),
  \qquad 
  (\mcN, \mcM) = ( (n_1,m_1),\ldots,(n_k,m_k)).
\]
That is, 
\begin{equation*}
\begin{aligned}
  \frac{1}{p!q!} 
  \sum_{D \in \mcL_{p, q}}
    \Gamma_D
  & =    
    \sum_{k=1}^\infty \frac{1}{k!} 
    \sum_{\substack{n_1,\ldots,n_k\geq 0 \\ m_1,\ldots,m_k\geq 0 \\ \textnormal{For each $\ell$: } n_\ell + m_\ell \geq 2}}
    \chi_{(\sum_{\ell} n_\ell = p)}
    \chi_{(\sum_{\ell} m_\ell = q)}
    \frac{1}{\prod_{\ell=1}^k n_\ell! m_\ell!}
  \\ & \quad \times 
    \idotsint 
    \ud X_{p} \ud Y_{q}
    \left[
    \prod_{\ell=1}^k 
    \sum_{G_\ell \in \mcC_{n_\ell,m_\ell}}
    \prod_{e\in G_\ell} g_e
    \right]
  \rho_t^{(\mcN, \mcM)}.
\end{aligned}
\end{equation*}
To bound this we use the tree-graph bound \cite{Ueltschi.2018}, see also \cite[Proposition 6.1]{Poghosyan.Ueltschi.2009}.
By assumption \Cref{eqn.tree.graph.condition} is satisfied and thus \cite{Ueltschi.2018}
\begin{equation}\label{eqn.apply.tree.graph}
  \abs{
  \sum_{G \in \mcC_{p,q}}
    \prod_{e\in G} g_e
  }
  \leq 
  C_{\textnormal{TG}}^{p+q}
  \sum_{T \in \mcT_{p,q}} \prod_{e\in T} |g_e|,
\end{equation}
where $\mcT_{p,q} \subset \mcG_{p,q}$ denotes the subset of trees. 
(To see this note that $\mcC_{p,q}$ (respectively $\mcT_{p,q}$) is the set of connected graphs (respectively trees) 
on $p+q$ vertices with the colours of the vertices just serving as a handy reminder of the edge-weights $g_e$.)
By moreover using the bound on the truncated correlation in \Cref{eqn.bound.truncated.correlation}
we conclude that 
\begin{equation}
\label{eqn.decompose.clusters.with.tree.graph}
\begin{aligned}
  \sum_{\substack{p,q\geq 0 \\ p+q \geq 2}}
  \frac{1}{p!q!} 
  \abs{\sum_{D \in \mcL_{p, q}}
    \Gamma_D}
  & \leq    
    \sum_{k=1}^\infty \frac{1}{k!} 
    \sum_{\substack{n_1,\ldots,n_k\geq 0 \\ m_1,\ldots,m_k\geq 0 \\ \textnormal{For each $\ell$: } n_\ell + m_\ell \geq 2}}
    \frac{1}{\prod_{\ell=1}^k n_\ell! m_\ell!}
    (C_{\textnormal{TG}} \gamma_\infty)^{\sum_\ell (n_\ell + m_\ell) - (k - 1)} 
    C_{\textnormal{TG}}^{k-1}
  \\ & \quad \times 
    \sum_{\substack{T_1,\ldots,T_k \\ T_\ell \in \mcT_{n_\ell,m_\ell}}}
    \sum_{A \in \mcA^{((n_1,m_1),\ldots,(n_k,m_k))}}
    \idotsint 
    \ud X_{p} \ud Y_{q}
    \left[
    \prod_{\ell=1}^k 
    \prod_{e\in T_\ell} |g_e|
    \right]
    \prod_{(\mu,\nu)\in A} \abs{\gamma_{\mu\nu}}.
\end{aligned}
\end{equation}
To do the integrations we note that the graph $\mathscr{T}$ with edges the union of ($g$-)edges in $T_1,\ldots,T_k$ and ($\gamma$-)edges in $A$ 
is a tree on all the $\sum_\ell n_\ell + \sum_\ell m_\ell$ many vertices.
One then integrates the coordinates one leaf at a time (meaning that the index of the corresponding coordinate is a leaf of the graph $\mathscr{T}$)
and removes a vertex from the graph after integrating over its corresponding coordinate.

To be more precise suppose that $\nu_0$ is a leaf of $\mathscr{T}$. 
Then the variable $z_{\nu_0}$
appears exactly once in the integrand. Either in a factor $g_{\mu\nu_0}$ (in which case the $z_{\nu_0}$-integral gives $\int |g| \leq I_g$ by the translation invariance) 
or in a factor $\gamma_{\mu\nu_0}$ (in which case the $z_{\nu_0}$-integral gives $\int |\gamma| \leq I_\gamma$ by the translation invariance).
The final integral gives $L^3$ by the translation invariance.
There are $k-1$ factors of $\gamma$ and $\sum_\ell(n_\ell + m_\ell - 1) = p+q-k$ factors of $g$.
Thus we get
\[
\begin{aligned}
  \sum_{\substack{p,q\geq 0 \\ p+q\geq 2}}
  \frac{1}{p!q!} 
  \abs{\sum_{D \in \mcL_{p, q}}
    \Gamma_D}
  & \leq    
    \sum_{k=1}^\infty \frac{1}{k!} 
    \sum_{\substack{n_1,\ldots,n_k\geq 0 \\ m_1,\ldots,m_k\geq 0 \\ \textnormal{For each $\ell$: } n_\ell + m_\ell \geq 2}}
    \frac{1}{\prod_{\ell=1}^k n_\ell! m_\ell!}
    \left[\prod_{\ell=1}^k 
            \# \mcT_{n_\ell,m_\ell}\right]
  \\ & \quad \times 
    \# \mcA^{((n_1,m_1),\ldots,(n_k,m_k))}
    (C_{\textnormal{TG}} I_g \gamma_\infty)^{\sum_\ell(n_\ell + m_\ell)-k}
    (C_{\textnormal{TG}}I_\gamma)^{k-1}
    C_{\textnormal{TG}}\gamma_\infty L^3.
\end{aligned}
\]
In \cite[Appendix D.5]{Giuliani.Mastropietro.ea.2021} it is shown that 
\[
  \# \mcA^{((n_1,m_1),\ldots,(n_k,m_k))}
  \leq k! C^{\sum_\ell (n_\ell + m_\ell)}.
\]
Moreover, $\mcT_{n,m} = (n+m)^{n+m-2} \leq C^{n+m} (n+m)!$ by Cayley's formula.
Finally, we may bound the binomial coefficients $\frac{(n+m)!}{n!m!}\leq 2^{n+m}$.
Thus 
\[
\begin{aligned}
  \sum_{\substack{p,q\geq 0 \\ p+q\geq 2}}
  \frac{1}{p!q!} 
  \abs{\sum_{D \in \mcL_{p, q}}
    \Gamma_D}
  & \leq  
  C L^3\gamma_\infty   
    \sum_{k=1}^\infty 
    \left[\sum_{\substack{n,m\geq 0 \\ n+m\geq 2}}
            \frac{(n+m)!}{n! m!}
            (C I_g \gamma_\infty)^{n + m - 1}\right]^k
    (C I_\gamma)^{k-1}
  \\
  & \leq 
  C L^3 \gamma_\infty 
  \sum_{k=1}^\infty \left[\sum_{\ell = 2}^\infty \ell (CI_g \gamma_\infty)^{\ell-1}\right]^k (C I_\gamma)^{k-1}
  \\ & \leq 
  C L^3 \gamma_\infty^2 I_g < \infty
\end{aligned}
\]
for $\gamma_\infty I_g$ and $\gamma_\infty I_g I_\gamma$ small enough.
This shows that 
$\sum_{p,q:  p+q \geq 2} \frac{1}{p!q!}\sum_{D \in \mcL_{p, q}}\Gamma_D$
is absolutely convergent under this assumption.
Next, we bound the $\Gamma^{n,m}$-sum for $n+m \geq 1$.

\subsection{Absolute convergence of the \texorpdfstring{$\Gamma^{n,m}$}{Gamma-nm}-sum}\label{sec.Gamma.nm.abs.conv}
In this section we prove that 
(for $n+m\geq 1$ and uniformly in $X_n, Y_m$)
\[
  \sum_{p,q\geq 0} \frac{1}{p!q!} 
  \abs{\sum_{D \in \mcL_{p, q}^{n, m}}
    \Gamma_D^{n,m}}
  \leq C_{n,m} \gamma_\infty^{n+m}
  < \infty
\]
if \Cref{eqn.tree.graph.condition} is satisfied and 
$\gamma_\infty I_g (1+I_\gamma)$ is sufficiently small.

We do the same splitting into clusters (connected components of $G$) as in \Cref{sec.Gamma.abs.conv} above.
There is however a slight complication:
One needs to keep track of in which clusters the external vertices lie.
This is exactly parametrized by the set $\Pi^{n,m}_\kappa$ (defined in \Cref{eqn.define.Pi.kappa}).
Denoting the sizes (number of internal vertices) of the clusters containing external vertices by $(n^*_\lambda, m^*_\lambda)$ 
and the sizes of clusters only containing internal vertices by $(n_\ell, m_\ell)$
and introducing $\mcC_{p,q}^{n,m} \subset \mcG_{p,q}^{n,m}$ as the subset of connected graphs 
(and similarly $\mcC_{B,W}^{B^*,W^*} \subset \mcG_{B,W}^{B^*,W^*}$, recall \Cref{def.diagrams})
we get
\begin{equation}\label{eqn.decompose.clusters.n+m.geq1}
\begin{aligned}
  & \frac{1}{p!q!} 
  \sum_{D \in \mcL_{p, q}^{n, m}}
    \Gamma_D^{n,m}
  \\ & \quad  =   
    \sum_{k=0}^\infty \frac{1}{k!} 
    \sum_{\kappa=1}^{n+m} \frac{1}{\kappa!}
    \sum_{(\mcB^*,\mcW^*)\in \Pi_\kappa^{n,m}}
    \sum_{\substack{n_{1}^*,\ldots,n_{\kappa}^*\geq 0 \\ m_{1}^*,\ldots,m_{\kappa}^*\geq 0}}
    \sum_{\substack{n_1,\ldots,n_k\geq 0 \\ m_1,\ldots,m_k\geq 0 \\ \textnormal{For each $\ell$: } n_\ell + m_\ell \geq 2}}
    \chi_{(\sum_{\ell} n_\ell + \sum_\lambda n^*_\lambda = p)}
    \chi_{(\sum_{\ell} m_\ell + \sum_\lambda m^*_\lambda = q)}
  \\ & \qquad \times
    \frac{1}{\prod_{\lambda=1}^\kappa n_{\lambda}^*! m_{\lambda}^*!}
    \frac{1}{\prod_{\ell=1}^k n_\ell! m_\ell!}
    \sum_{G^*_\lambda \in \mcC_{n^*_\lambda,m^*_\lambda}^{B_\lambda^*, W_\lambda^*}}
    \sum_{G_\ell \in \mcC_{n_\ell,m_\ell}}
    \idotsint 
    \ud X_{[n+1,n+p]} \ud Y_{[m+1,m+q]}
  \\ & \qquad \times 
    \left[
    \prod_{\lambda=1}^\kappa \prod_{e\in G^*_\lambda} g_e
    \right] 
    \times 
    \left[
    \prod_{\ell=1}^k \prod_{e\in G_\ell} g_e
    \right]
  \\ & \qquad \times 
    \left[
    \sum_{\substack{\pi \in \mcS_{n+p} \\ \tau \in \mcS_{m+q}}}
    (-1)^\pi (-1)^\tau 
    \chi_{( (\pi, \tau, \cup_\lambda G_\lambda^* \cup \cup_\ell G_\ell) \textnormal{ linked})}
    \prod_{i=1}^{n+p} \gamma_{\uparrow}(x_i - x_{\pi(i)})
    \prod_{j=1}^{m+q} \gamma_{\downarrow}(y_j - y_{\tau(j)})
    \right].
\end{aligned}
\end{equation}
For $k=0$ the $n_1,m_1,\ldots,n_k,m_k$-sum should be interpreted as an empty product, i.e. as a factor $1$.
Similarly for $p=0$ and/or $q=0$ the empty product of integrals should be interpreted as a factor $1$.

The last line in \Cref{eqn.decompose.clusters.n+m.geq1} is the truncated correlation
\[
\rho_t^{(\mcB^* + \mcN^*, \mcW^* + \mcM^*) \oplus (\mcN, \mcM)},
\]
where 
\[
    \mcN^* = (n^*_1, \ldots, n^*_\kappa), 
    \qquad 
    \mcN = (n_1, \ldots, n_k),
    \qquad 
    \mcM^* = (m^*_1, \ldots, m^*_\kappa),
    \qquad 
    \mcM = (m_1,\ldots,m_k)
\]
and $\oplus$ means concatenation of vectors, 
i.e. 
\begin{multline*}
    (\mcB^* + \mcN^*, \mcW^* + \mcM^*) \oplus (\mcN,\mcM)
    \\ = 
    ( (B_1^* + n_1^*, W_1^* + m_1^*), \ldots, (B_\kappa^* + n_\kappa^*, W_\kappa^* + m_\kappa^*), 
    (n_1, m_1), \ldots, (n_k, m_k) ),    
\end{multline*}
where we abused notation slightly and wrote $B_1^* + n_1^*$ for the union of the vertices $B_1^*$ and the $n_1^*$ internal black vertices of the graph $G_1^*$.
(Similarly for the other terms.)

We use as in \Cref{sec.Gamma.abs.conv} the tree-graph bound and the bound on the truncated correlation in \Cref{eqn.bound.truncated.correlation}.
For the clusters with external vertices we add $0$-weights to the disallowed edges as in \cite[Section 3.1.3]{Lauritsen.Seiringer.2024},
i.e. for $G\in \mcC_{p,q}^{n,m}$ define 
\[
  \tilde g_e = \begin{cases}
  0 & e = (i,j) \textnormal{ with } i,j \textnormal{ external vertices}
  \\
  g_e & \textnormal{otherwise}.
  \end{cases}
\]
Then we may readily apply the tree-graph bound \cite{Ueltschi.2018} with edge-weights $\tilde g_e$:
\[
  \abs{\sum_{G\in \mcC_{p,q}^{n,m}} \prod_{e\in G} g_e}
  =
  \abs{\sum_{G\in \mcC_{p+n,q+m}} \prod_{e\in G} \tilde g_e}
  \leq 
  C_{\textnormal{TG}}^{p+q+n+m}
  \sum_{T\in \mcT_{p+n, q+m}} \prod_{e\in T} \abs{\tilde g_e}
  =
  C_{\textnormal{TG}}^{p+q+n+m}
  \sum_{T\in \mcT_{p, q}^{n,m}} \prod_{e\in T} \abs{g_e},
\]
where 
$\mcT_{p,q} \subset \mcG_{p,q}$ and $\mcT_{p,q}^{n,m} \subset \mcC_{p,q}^{n,m}$ denotes the subsets of trees.
Thus
\begin{equation}\label{eqn.bound.abs.Gamma.nm.initial}
\begin{aligned}
  & 
  \sum_{p,q\geq 0} 
  \frac{1}{p!q!} 
  \abs{\sum_{D \in \mcL_{p, q}^{n, m}}
    \Gamma_D^{n,m}}
  \\ 
  & 
  \quad 
    \leq  
    \sum_{k=0}^\infty \frac{1}{k!} 
    \sum_{\kappa=1}^{n+m} \frac{1}{\kappa!}
    \sum_{(\mcB^*,\mcW^*)\in \Pi_\kappa^{n,m}}
    \sum_{\substack{n_{1}^*,\ldots,n_{\kappa}^*\geq 0 \\ m_{1}^*,\ldots,m_{\kappa}^*\geq 0}}
    \sum_{\substack{n_1,\ldots,n_k\geq 0 \\ m_1,\ldots,m_k\geq 0 \\ \textnormal{For each $\ell$: } n_\ell + m_\ell \geq 2}}
    \frac{1}{\prod_{i=1}^\ell n_{i}^*! m_{i}^*!}
    \frac{1}{\prod_{\ell=1}^k n_\ell! m_\ell!}
  \\ & 
  \qquad 
  \times
    \sum_{A \in \mcA^{(\mcB^* + \mcN^*, \mcW^* + \mcM^*) \oplus (\mcN,\mcM)}}
    \sum_{\substack{T^*_1,\ldots,T^*_\kappa \\ T^*_\lambda \in \mcT_{n^*_\lambda, m^*_\lambda}^{B_\lambda^*, W_\lambda^*}}}
    \sum_{\substack{T_1,\ldots,T_k \\ T_\ell \in \mcT_{n_\ell,m_\ell}}}
  \\ & \qquad \times 
    \idotsint 
    \ud X_{\left[n+1,n + \sum_\lambda n^*_\lambda + \sum_\ell n_\ell\right]} 
    \ud Y_{\left[m+1,m + \sum_\lambda m^*_\lambda + \sum_\ell m_\ell\right]} 
    \Bigg[
    \prod_{\lambda=1}^\kappa \prod_{e\in T^*_\lambda} |g_e|
    \prod_{\ell=1}^k \prod_{e\in T_\ell} |g_e|
    \prod_{(\mu,\nu) \in A} \abs{\gamma_{\mu\nu}}
    \Bigg]
  \\ & \qquad \times 
    (C_{\textnormal{TG}}\gamma_\infty)^{\sum_\lambda (n^*_\lambda + m^*_\lambda) + \sum_\ell (n_\ell + m_\ell) + n + m - (k + \kappa - 1)}
    C_{\textnormal{TG}}^{k+\kappa - 1}.
\end{aligned}
\end{equation}
To do the integrations we bound some $g$- and $\gamma$-factors pointwise. 
Recall first, that there are $\kappa$ clusters with external vertices.
We split the anchored tree into pieces according to these clusters as follows.

We may view the anchored tree $A$ as a tree on the set of clusters. 
If $\kappa = 1$ set $A_1 = A$. Otherwise 
iteratively pick a $\gamma$-edge on the path in $A$ between any two clusters with external vertices 
and bound it by 
\[
  \abs{\gamma_\sigma(z)} = \abs{\sum_{k\in \frac{2\pi}{L}\Z^3} \hat \gamma_\sigma(k) e^{ikz}}\leq \gamma_\infty
\] 
and remove it from $A$. This cuts the anchored tree $A$ into pieces.
Doing this $\kappa-1$ many times we get $\kappa$ anchored trees $A_1,\ldots,A_\kappa$
with each exactly one cluster with external vertices. 
That is,
\[
  \prod_{(\mu,\nu) \in A} |\gamma_{\mu\nu}| 
  \leq \gamma_\infty^{\kappa-1} \prod_{\lambda=1}^\kappa \prod_{(\mu,\nu)\in A_\lambda} |\gamma_{\mu\nu}|.
\]
Next, in each cluster with external vertices, say with label $\lambda_0$, we do a similar procedure 
of splitting the cluster into pieces according to the external vertices.

In the cluster $\lambda_0$ there are $\# B_{\lambda_0}^* + \# W_{\lambda_0}^* \geq 1$ external vertices. 
If $\# B_{\lambda_0}^* + \# W_{\lambda_0}^* = 1$ set $T_{\lambda_0,1}^* = T_{\lambda_0}^*$. 
Otherwise iteratively pick a $g$-edge on the path in $T_{\lambda_0}^*$ between any two external vertices
and bound it by
\[
  \abs{g_e} = \abs{f_e^2 - 1} \leq \max \{f_e^2, 1\} \leq C_{\textnormal{TG}}^2
\] 
using \Cref{eqn.tree.graph.condition} for $q=2$.
Remove the edge $e$ from $T_{\lambda_0}^*$. This cuts the tree $T_{\lambda_0}^*$ into pieces.
Doing this $\# B_{\lambda_0}^* + \# W_{\lambda_0}^* - 1$ many times we get $\# B_{\lambda_0}^* + \# W_{\lambda_0}^*$ 
trees $T_{\lambda_0, 1}^*, \ldots, T_{\lambda_0, \# B_{\lambda_0}^* + \# W_{\lambda_0}^*}^*$
with each exactly one external vertex. That is,
\[
  \prod_{e\in T_{\lambda_0}^*} |g_e|
  \leq C_{\textnormal{TG}}^{2(\#B_{\lambda_0}^* + \# W_{\lambda_0}^* - 1)}
  \prod_{\nu = 1}^{\# B_{\lambda_0}^*  + \# W_{\lambda_0}^*} \prod_{e\in T_{\lambda_0, \nu}^*} |g_e|.
\]
We do this procedure for all the $\kappa$ many clusters with external vertices. 
Then the graph $\mathscr{T}$ with edges the union of all ($g$- or $\gamma$-)edges in $T_{\lambda, \nu}^*, T_\ell, A_\lambda$  
(for $\lambda \in \{1,\ldots,\kappa\}$, $\ell \in \{1,\ldots,k\}$ and $\nu \in \{1,\ldots,\# B_\lambda^* + \# W_\lambda^*-1\}$)
is a 
forest (disjoint union of trees) on the set of vertices 
$V_{n + \sum_\lambda n_\lambda^* + \sum_\ell n_\ell, m + \sum_\lambda m^*_\lambda + \sum_\ell m_\ell}$ 
with each connected component (tree) having exactly one external vertex.
Moreover, we have the bound
\begin{equation}\label{eqn.cut.trees}
\begin{aligned}
& \idotsint 
  \ud X_{\left[n+1,n + \sum_\lambda n^*_\lambda + \sum_\ell n_\ell\right]} 
  \ud Y_{\left[m+1,m + \sum_\lambda m^*_\lambda + \sum_\ell m_\ell\right]} 
    \Bigg[
    \prod_{\lambda = 1}^\kappa \prod_{e\in T^*_\lambda} |g_e|
    \prod_{\ell = 1}^k \prod_{e\in T_\ell} |g_e|
    \prod_{(\mu,\nu) \in A} \abs{\gamma_{\mu\nu}}
    \Bigg]
\\ & \quad 
  \leq
  C_{\textnormal{TG}}^{2(n+m-\kappa)}
  \gamma_\infty^{\kappa - 1}
  \left[
    \prod_{\lambda=1}^\kappa \prod_{\nu=1}^{\# B_\lambda^* + \# W_\lambda^*} 
    \idotsint \prod_{e\in T_{\lambda, \nu}^*} |g_e| \prod_{(\mu,\nu)\in A_\lambda} |\gamma_{\mu\nu}| \prod_{\ell: T_\ell \sim A_\lambda} \prod_{e\in T_\ell}|g_e|
  \right],
\end{aligned}
\end{equation}
where $T_\ell \sim A_\lambda$ means that $T_\ell$ and $A_\lambda$ share a vertex. 
(Equivalently they are part of the same connected component of $\mathscr{T}$.)

Since each connected component of $\mathscr{T}$ is a tree we may do the integrations 
one leaf at a time exactly as for the $\Gamma$-sum in \Cref{sec.Gamma.abs.conv} above.
To bound the value we count the number of $\gamma$- and $g$-factors that are left.

The number of $\gamma$-integrations is exactly the number of $\gamma$-factors.
There are $k+\kappa$ many clusters, so $A$ has $k+\kappa - 1$ many edges. 
In constructing $A_1,\ldots,A_\kappa$ we cut $\kappa-1$ many edges, thus there is 
$k$ many $\gamma$-factors left and so there are $k$ many $\gamma$-integrations in \Cref{eqn.cut.trees}.
The remaining $\sum_\lambda (n_\lambda^* + m_\lambda^*) + \sum_\ell (n_\ell + m_\ell) - k$ integrations
are of $g$-factors. 
The integrals may be bounded by $\int \abs{\gamma} \leq I_\gamma$ and $\int \abs{g} \leq I_g$ as in \Cref{sec.Gamma.abs.conv}.
Moreover, since each connected component of $\mathscr{T}$ has one external vertex, which is not integrated over, 
there are no volume factors from the last integrations in any of the connected components of $\mathscr{T}$.
That is,
\begin{equation*}
\begin{aligned}
& \idotsint 
  \ud X_{\left[n+1,n + \sum_\lambda n^*_\lambda + \sum_\ell n_\ell\right]} 
  \ud Y_{\left[m+1,m + \sum_\lambda m^*_\lambda + \sum_\ell m_\ell\right]} 
    \Bigg[
    \prod_{\lambda = 1}^\kappa \prod_{e\in T^*_\lambda} |g_e|
    \prod_{\ell = 1}^k \prod_{e\in T_\ell} |g_e|
    \prod_{(\mu,\nu) \in A} \abs{\gamma_{\mu\nu}}
    \Bigg]
\\ & \quad 
  \leq
  C_{\textnormal{TG}}^{2(n+m-\kappa)}
  \gamma_\infty^{\kappa - 1}
  I_g^{\sum_\lambda (n_\lambda^* + m_\lambda^*) + \sum_\ell (n_\ell + m_\ell) - k}
  I_\gamma^{k}.
\end{aligned}
\end{equation*}
We use this to bound the integrations in \Cref{eqn.bound.abs.Gamma.nm.initial}.
Additionally we need to bound the number of (anchored) tree. 
In \cite[Appendix D.5]{Giuliani.Mastropietro.ea.2021} it is shown that 
\[
  \# \mcA^{(\mcB^* + \mcN^*, \mcW^* + \mcM^*) \oplus (\mcN,\mcM)}
  \leq (k+\kappa)! C^{n + m + \sum_\lambda (n^*_\lambda + m^*_\lambda) + \sum_\ell (n_\ell + m_\ell)},
\]
since we have $k+\kappa$ many clusters and $n + m + \sum_\lambda (n^*_\lambda + m^*_\lambda) + \sum_\ell (n_\ell + m_\ell)$ 
many vertices in total.
Moreover, 
$\# \mcT_{p,q}^{n,m} \leq \# \mcT_{p+n,q+m}  = (p + q+n+m)^{p + q+n+m - 2} \leq (p+q+n+m)! C^{p+q+n+m}$
by Cayley's formula as in \Cref{sec.Gamma.abs.conv}.
These bounds together with \Cref{eqn.bound.abs.Gamma.nm.initial} then gives
\begin{equation}\label{eqn.bound.Gamma.nm.pre.binomial}
\begin{aligned}
  & 
  \sum_{p,q\geq 0} 
  \frac{1}{p!q!} 
  \abs{\sum_{D \in \mcL_{p, q}^{n, m}}
    \Gamma_D^{n,m}}
  \\ 
  & 
  \quad 
    \leq  
    (C\gamma_\infty)^{n+m}
    \sum_{k=0}^\infty 
    \sum_{\kappa=1}^{n+m} 
    \frac{(k+\kappa)!}{k!\kappa!} 
    \sum_{(\mcB^*,\mcW^*)\in \Pi_\kappa^{n,m}}
    \sum_{\substack{n_{1}^*,\ldots,n_{\kappa}^*\geq 0 \\ m_{1}^*,\ldots,m_{\kappa}^*\geq 0}}
    \sum_{\substack{n_1,\ldots,n_k\geq 0 \\ m_1,\ldots,m_k\geq 0 \\ \textnormal{For each $\ell$: } n_\ell + m_\ell \geq 2}}
  \\ & 
  \qquad 
  \times
    \left[\prod_{\lambda=1}^\kappa 
      \frac{(n^*_\lambda + \#B_\lambda^* + m^*_\lambda + \# W_\lambda^*)!}{n^*_\lambda! m^*_\lambda!}
      \right]
    \left[\prod_{\ell=1}^k 
      \frac{(n_\ell + m_\ell)!}{n_\ell!m_\ell!}\right]
    (C I_g \gamma_\infty)^{\sum_\lambda (n^*_\lambda + m^*_\lambda) + \sum_{\ell}(n_\ell + m_\ell - 1)}
    (CI_\gamma)^k.
\end{aligned}
\end{equation}
Multinomial coefficients may be bounded as $\frac{(p_1+\ldots + p_k)!}{p_1!\cdots p_k!} \leq k^{p_1+\ldots+p_k}$. Moreover, 
$\# B_\lambda^* \leq n$ and $\# W_\lambda^*\leq m$.
Thus we may bound 
\[
  (n^*_\lambda + \# B_\lambda  + m^*_\lambda + \# W_\lambda)!
  \leq (n^*_\lambda + m^*_\lambda + n + m)!
  \leq 4^{n^*_\lambda + m^*_\lambda + n + m} n! m! n^*_\lambda! m^*_\lambda!.
\]
We conclude the bound 
\begin{equation}\label{eqn.final.bound.Gamma.nm}
\begin{aligned}
  & 
  \sum_{p,q\geq 0} 
  \frac{1}{p!q!} 
  \abs{\sum_{D \in \mcL_{p, q}^{n, m}}
    \Gamma_D^{n,m}}
  \\ 
  & 
  \quad 
    \leq  
    (C\gamma_\infty)^{n+m}
    \sum_{k=0}^\infty 
    \sum_{\kappa=1}^{n+m} 
    2^{k + \kappa}
    \left[\sum_{n^*_0, m^*_0 \geq 0}
      C_{n,m} (C I_g \gamma_\infty)^{n^*_0+ m^*_0}\right]^\kappa
    \left[
    C I_\gamma
    \sum_{\substack{n_0, m_0 \geq 0 \\ n_0 + m_0 \geq 2}}
      (C I_g \gamma_\infty)^{n_0 + m_0 - 1}
      \right]^k.
\end{aligned}
\end{equation}
For some $c_{n,m} > 0$ we have that if $\gamma_\infty I_g (1+I_\gamma) < c_{n,m}$ the sums are convergent and we get 
\[
  \sum_{p,q\geq 0} 
  \frac{1}{p!q!} 
  \abs{\sum_{D \in \mcL_{p, q}^{n, m}}
    \Gamma_D^{n,m}}
    \leq  
    C_{n,m}\gamma_\infty^{n+m}
    < \infty.
\]
This shows the desired. 
We conclude the proof of \Cref{lemma.gaudin.general} for the case $S=2$.

\begin{remark}[Higher spin]\label{rmk.higher.spin.multinomial}
For the case of higher spin $S\geq 3$, the computations are essentially the same. 
\end{remark}

\noindent
For later use we define for all diagrams some values characterising their sizes.
\begin{defn}\label{def.k.ng}
Let $D\in \mcL^{n,m}_{p,q}$. 
Define the number $k=k(D)$ as the number of clusters entirely within 
internal vertices (i.e. the same $k$ as in the computations above)
and $\kappa = \kappa(D)$ as the number of clusters containing at least one external vertex (i.e. the same $\kappa$ as in the computations above).
Define then $\nu^* = \nu^*(D)$ and $\nu = \nu(D)$ as 
\[
    \nu^* = \sum_{\lambda=1}^\kappa (n_\lambda^* + m_\lambda^*),
    \qquad 
    \nu = \sum_{\ell=1}^k (n_\ell + m_\ell) - 2k,
\]
where $n^*_\lambda, m^*_\lambda, n_\ell, m_\ell$ are the sizes of the different clusters exactly as in the computations above.
(Then $\nu + \nu^{*} + 2k= p+q$.)
\end{defn}

\noindent
For a diagram $D$ the number $\nu + \nu^*$ is the ``number of added vertices'' in the following sense.
A diagram with $n+m$ external vertices and $k$ clusters entirely within 
internal vertices has at least $n+m+2k$ many vertices, 
since each cluster (with only internal vertices) has at least $2$ vertices.
Then $\nu + \nu^*$ is the number of vertices a diagram has more than this minimal number.

Note that in the special case of consideration with the scattering functions $f_s, f_p$ 
and the one-particle density matrices $\gamma_{N_\sigma}^{(1)}$ we have
\[
\gamma_\infty \leq \rho, 
\qquad 
I_g \leq Cab^2, 
\qquad 
I_\gamma \leq C s (\log N)^3
\]
by \Cref{eqn.bound.int.g,eqn.derivative.lebesgue.constants}, see also the proof of \Cref{thm.gaudin}.
Then, by following the arguments above (see in particular \Cref{eqn.final.bound.Gamma.nm,eqn.bound.Gamma.nm.pre.binomial}), 
we have (for $p + q =2k_0 + \nu_0$)
\begin{equation}\label{eqn.k.ng.diagrams}
    \frac{1}{p!q!} 
    \abs{
    \sum_{\substack{D \in \mcL_{p,q}^{n,m} 
        \\ k(D)= k_0 
        \\ \nu(D) + \nu^*(D) = \nu_0}}
        \Gamma_D^{n,m}
    }
    \leq C_{n,m}\rho^{n+m} 
    (Cab^2\rho)^{\nu_0 + k_0} (Cs (\log N)^3)^{k_0}
\end{equation}
for any $n,m$ with $n+m\geq 1$.
We think of $s$ as $s \sim (a^3\rho)^{-1/3 + \eps}$ for some small $\eps > 0$.
Thus increasing $\nu_0$ by $1$ we decrease the size of the diagram by $(a^3\rho)^{1/3}$, 
and increasing $k_0$ by $1$ we decrease the size of the diagram by $(a^3\rho)^{\eps}$.
(Recall that $b=\rho^{-1/3}$.)

\section{Energy of the trial state}\label{sec.energy}
In this section we use the formulas in \Cref{eqn.thm.gaudin.main} to calculate the energy in \Cref{eqn.energy.compute}.
We will refer to a term in \Cref{eqn.energy.compute} where $\rho^{(n,m)}_{\textnormal{Jas}}$ appears as 
a $(n,m)$-type term.

\subsection{\texorpdfstring{$2$}{2}-body terms}
In this section we consider the terms in \Cref{eqn.energy.compute} 
where a two-particle density ($\rho_{\textnormal{Jas}}^{(n,m)}$ with $n+m=2$) appears.
We consider first the term with $m=n=1$.

\subsubsection{\texorpdfstring{$(1,1)$}{(1,1)}-type terms}
We consider the term 
\begin{equation}\label{eqn.type.(1.1).contribution}
    2 \iint \rho_{\textnormal{Jas}}^{(1,1)}
    \left[
    \frac{\abs{\nabla f_s(x_1-y_1)}^2}{f_s(x_1-y_1)^2}
    + \frac{1}{2}v(x_1-y_1)
    \right]
    \ud x_1 \ud y_1.
\end{equation}
The formula in \Cref{eqn.thm.gaudin.main} reads 
for $\rho_{\textnormal{Jas}}^{(1,1)}$ as follows.
\begin{equation}\label{eqn.compute.rho.(1.1)}
\begin{aligned}
    \rho^{(1,1)}_{\textnormal{Jas}}(x_1,y_1) 
    & = f_s(x_1-y_1)^2
    \left[ 
        \rho_{\textnormal{Jas}}^{(1,0)} \rho_{\textnormal{Jas}}^{(0,1)}
        +
        \sum_{\substack{p,q \geq 0}}
        \frac{1}{p!q!} \sum_{D\in \mcL_{p,q}^{1,1}} \Gamma_{D}^{1,1}
    \right]
    \\
    & = 
    f_s(x_1-y_1)^2 
    \left[ 
        \rho_\uparrow \rho_\downarrow
        + \sum_{\substack{p,q \geq 0 \\ p+q\geq 1}}
        \frac{1}{p!q!} \sum_{D\in \mcL_{p,q}^{1,1}} \Gamma_{D}^{1,1}
    \right]
\end{aligned}
\end{equation}
since $\mcL_{p,q}^{1,1} = \varnothing$ for $p=q=0$.
The second summand is an error term. We bound it as follows.

\begin{lemma}\label{lem.bound.error.rho.(1.1)}
    There exists a constant $c>0$ such that if $s a b^2 \rho (\log N)^3 < c$,
    then for any integer $K$ there exists a constant $C_K > 0$ such that
    \[
        \sum_{\substack{p,q \geq 0 \\ p+q\geq 1}}
        \frac{1}{p!q!} 
        \abs{\sum_{D\in \mcL_{p,q}^{1,1}} \Gamma_{D}^{1,1}}
        \leq 
        C_K a b^2 \rho^3
        + C \rho^2 (Csab^2 \rho (\log N)^3)^{K+1}
        + C s a^3 \rho^3 \log (b/a)(\log N)^3.
    \]
\end{lemma}

\noindent
We give the proof at the end of this section.

Using \Cref{eqn.compute.rho.(1.1)} and \Cref{lem.bound.error.rho.(1.1)} we get 
for any integer $K$
\[
\begin{aligned}
\eqref{eqn.type.(1.1).contribution}
& = 2 L^3 \int \left(|\nabla f_s|^2 + \frac{1}{2} v f_s^2\right) \ud x
\\ & \quad \times 
\bigl[
\rho_\uparrow \rho_\downarrow 
    + O_K\left(ab^2\rho^3\right)
    + O_K\left(\rho^2 (sab^2 \rho(\log N)^3)^{K+1}\right)
    + O\left(sa^3 \log (b/a) (\log N)^3\right)
\bigr].    
\end{aligned}
\]
By \Cref{def.scattering.length} we have 
\[
\int \left(|\nabla f_s|^2 + \frac{1}{2} v f_s^2\right) \ud x
\leq \frac{1}{(1 - a/b)^2} 
\int \left(|\nabla f_{s0}|^2 + \frac{1}{2} v f_{s0}^2\right) \ud x
= \frac{4\pi a}{(1 - a/b)^2} 
= 4\pi a + O(a^2/b).
\]
We conclude that 
\begin{equation}\label{eqn.contribution.final.(1.1).terms}
\begin{aligned}
\eqref{eqn.type.(1.1).contribution}
& \leq L^3 8\pi a \rho_\uparrow \rho_\downarrow 
 + O(L^3 a^2b^{-1}\rho^2)
 + O_K(L^3 a^2 b^2 \rho^3)
 + O_K\left(L^3 a\rho^2 (sab^2 \rho (\log N)^3)^{K+1}\right)
\\ & \quad 
 + O(L^3 s a^3 \rho^3 \log(b/a) (\log N)^3).
\end{aligned}
\end{equation}
Finally, we  give the

\begin{proof}[{Proof of \Cref{lem.bound.error.rho.(1.1)}}]
We split the diagrams into three groups using the numbers $\nu^*, \nu$ and $k$ from \Cref{def.k.ng}: 
\begin{enumerate}[(A)]
    \item Diagrams with $\nu+ \nu^*\geq 1$,
    \item Diagrams with $\nu + \nu^* = 0$,
    \begin{enumerate}[(B1)]
        \item at least one $p$-wave $g$-factor.
        \item only $s$-wave $g$-factors.
    \end{enumerate}
\end{enumerate}

\begin{remark}
The diagrams of types (A) and (B1) are those for which the bound in \Cref{eqn.k.ng.diagrams} 
is good enough to show that these diagrams give contributions to the energy density $\leq C a^2\rho^{7/3}$.
Naively using the bound in \Cref{eqn.k.ng.diagrams} for the diagrams of type (B2) we only get that these are bounded by $\rho^2 (a^3\rho)^{\eps}$  
with $b=\rho^{-1/3}$ and $s$ chosen as described immediately after \Cref{eqn.k.ng.diagrams}.
We will calculate the value of all the (infinitely many) diagrams of type (B2) below 
and use this exact calculation for all diagrams up to some arbitrary high order. 
This is an essential step in proving the ``almost optimal'' error bound in \Cref{thm.main}.
It is similar to the approach in \cite{Basti.Cenatiempo.ea.2023a} for the dilute Bose gas.
\end{remark}

\noindent
The contribution of all diagrams of type (A) (with $\nu+ \nu^*\geq 1$) is $\leq Ca b^2\rho^{3}$
by \Cref{eqn.k.ng.diagrams} if $s a b^2 (\log N)^3$ is sufficiently small (recall \Cref{thm.gaudin}).
For diagrams of type (B) note that we have $k\geq 1$, since any summand $p,q$ has $p+q\geq 1$.
Moreover, for diagrams of type (B1), at least one factor $\int |g_s| \leq Cab^2$ should be replaced by 
$\int |g_p| \leq C a^3\log b/a$ (recall the bounds in \Cref{eqn.bound.int.g}).
Thus, again by \Cref{eqn.k.ng.diagrams}, 
we may bound the size of all diagrams of type (B1) by 
$C s a^3 \rho^3 \log(b/a) (\log N)^3$.
More precisely we have 
\[ 
    \sum_{\substack{p,q\geq 0 \\ p+q\geq 1}} \frac{1}{p!q!} \abs{
    \sum_{\substack{ D \in \mcL_{p,q}^{1,1} \\ D \textnormal{ of type (A)}}}
    \Gamma_D^{1,1}
    }
    \leq C a b^2 \rho^3,
    \qquad 
    \sum_{\substack{p,q\geq 0 \\ p+q\geq 1}} \frac{1}{p!q!} \abs{
    \sum_{\substack{ D \in \mcL_{p,q}^{1,1} \\ D \textnormal{ of type (B1)}}}
    \Gamma_D^{1,1}
    }
    \leq C s a^3 \rho^3 \log (b/a) (\log N)^3
\]
if $sab^2 \rho(\log N)^3$ is sufficiently small.
It remains to consider the diagrams of type (B2), where $\nu + \nu^* = 0$ and only $s$-wave $g$-factors appear.
These diagrams have $g$-graph as in \Cref{fig.B2.general}.
Note that in particular $p=q=k(D)$ for any such diagram.

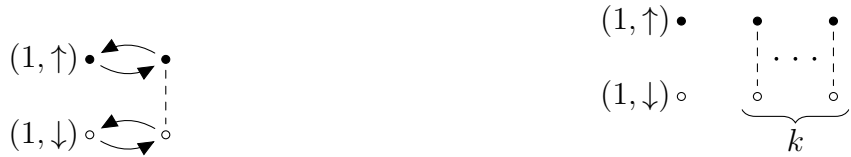
\begin{figure}[htb]
    \centering
\begin{subfigure}[t]{0.4\textwidth}
\centering
\begin{tikzpicture}[line cap=round,line join=round,>=triangle 45,x=1.0cm,y=1.0cm]
\node (1) at (0,1) {};
\node (2) at (0,0) {};
\node (3) at (1,1) {};
\node (4) at (1,0) {};
\draw[dashed] (3) -- (4);
\foreach \i in {1,3} \draw[fill] (\i) circle [radius=1.5pt];
\foreach \i in {2,4} \draw (\i) circle [radius=1.5pt];
\node[anchor = east] at (1) {$(1,\uparrow)$};
\node[anchor = east] at (2) {$(1,\downarrow)$};
\draw[->] (1) to[bend right] (3);
\draw[->] (2) to[bend right] (4);
\draw[->] (3) to[bend right] (1);
\draw[->] (4) to[bend right] (2);
\end{tikzpicture}
\caption{Diagram $D_\textnormal{small}$ of smallest size}
\label{fig.B2.small}
\end{subfigure}
\hspace{3em}
\begin{subfigure}[t]{0.4\textwidth}
\centering
\begin{tikzpicture}[line cap=round,line join=round,>=triangle 45,x=1.0cm,y=1.0cm]
\node (1) at (0,1) {};
\node (2) at (0,0) {};
\node (3) at (1,1) {};
\node (4) at (1,0) {};
\node (5) at (2,1) {};
\node (6) at (2,0) {};
\draw[dashed] (3) -- (4);
\draw[dashed] (5) -- (6);
\foreach \i in {1,3,5} \draw[fill] (\i) circle [radius=1.5pt];
\foreach \i in {2,4,6} \draw (\i) circle [radius=1.5pt];
\node[anchor = east] at (1) {$(1,\uparrow)$};
\node[anchor = east] at (2) {$(1,\downarrow)$};
\foreach \i in {1,...,3} \draw[fill] (1 + \i/4, 0.5) circle [radius=0.5pt];
\draw[decoration={brace,raise=5pt,amplitude=5pt,mirror},decorate] (0.8,0) -- node[below=8pt] {$k$} (2.2,0);
\end{tikzpicture}
\caption{Graph $G$ of general diagram, $k\geq 1$.}
\label{fig.B2.general}
\end{subfigure}
    \caption{Diagrams of type (B2).
    In (\protect\subref{fig.B2.general}) only the $g$-graph $G$ is drawn. 
    The relevant diagrams $(\pi,\tau,G)$ have $\pi,\tau$ such that the diagrams are linked.}
\end{figure}

We now evaluate all these diagrams.
We give an example calculation of the (unique) diagram of smallest size, and then do the computation in full generality.
The diagram of smallest size is the diagram in \Cref{fig.B2.small}. Its value is
\[
\begin{aligned}
\Gamma_{D_\textnormal{small}}^{1,1}
    & = \iint \gamma_{N_\uparrow}^{(1)}(x_1;x_2)\gamma_{N_\uparrow}^{(1)}(x_2;x_1)
    \gamma_{N_\downarrow}^{(1)}(y_1;y_2)\gamma_{N_\downarrow}^{(1)}(y_1;y_2)
    g_s(x_2 - y_2) \ud x_2 \ud y_2 
    \\
    & = 
    \frac{1}{L^{12}}
    \sum_{\substack{ k_1^\uparrow, k_2^\uparrow \in P_F^\uparrow 
    }}
    \sum_{k_1^\downarrow, k_2^\downarrow \in P_F^\downarrow}
    \iint e^{ik_1^\uparrow(x_1-x_2)} e^{ik_2^\uparrow(x_2-x_1)}
    e^{ik_1^\downarrow(y_1 - y_2)} e^{ik_2^\downarrow(y_2-y_1)}
    g_s(x_2 - y_2) \ud x_2 \ud y_2
    \\
    & = 
    \frac{1}{L^{12}}
    \sum_{\substack{ k_1^\uparrow, k_2^\uparrow \in P_F^\uparrow 
    }}
    \sum_{k_1^\downarrow, k_2^\downarrow \in P_F^\downarrow}
    e^{i(k_1^\uparrow - k_2^\uparrow) x_1}
    e^{i(k_1^\downarrow - k_2^\downarrow)y_1}
    \\ & \quad \times 
    \int \ud x_2 
    \left[ 
    e^{i(k_2^\uparrow - k_1^\uparrow + k_2^\downarrow - k_1^\downarrow)x_2}
    \int \ud y_2 
    \left(
    g_s(x_2-y_2) e^{-i(k_2^\downarrow - k_1^\downarrow)(x_2-y_2)}
    \right)
    \right]
    \\
    & = 
    \frac{1}{L^{12}}
    \sum_{\substack{ k_1^\uparrow, k_2^\uparrow \in P_F^\uparrow 
    }}
    \sum_{k_1^\downarrow, k_2^\downarrow \in P_F^\downarrow}
    e^{i(k_1^\uparrow - k_2^\uparrow) x_1}
    e^{i(k_1^\downarrow - k_2^\downarrow)y_1}
    L^3\chi_{(k_2^\uparrow - k_1^\uparrow = k_1^\downarrow - k_2^\downarrow)}
    L^3\hat g_s(k_2^\downarrow - k_1^\downarrow)
\end{aligned}
\]
where 
$\hat g_s(k) = L^{-3}\int g_s(x) e^{-ikx} \ud x$ denotes the Fourier transform
and we used the translation invariance to evaluate the $g_s$-integral.
We have the bound (recall \Cref{eqn.bound.int.g})
\[
L^3\abs{\hat g_s(k)} \leq \int |g(x)| \ud x 
\leq C a b^2.
\]
The characteristic function $\chi_{(k_2^\uparrow - k_1^\uparrow = k_1^\downarrow - k_2^\downarrow)}$ effectively kills one of the four $k^\sigma_j$-sums. 
The remaining $k_j^\sigma$-sums have at most $N_\sigma \leq N$ many summands.
We conclude the bound (uniformly in $x_1,y_1$)
\[
\abs{\Gamma_{D_\textnormal{small}}^{1,1}} \leq C a b^2 \rho^3.
\]
For the general diagram in \Cref{fig.B2.general} we may use the same method.
We then have 
\[
\begin{aligned}
\Gamma_{D}^{1,1}
&  = 
\frac{1}{L^{6+6k}} 
\sum_{k_1^\uparrow, \ldots, k_{k+1}^\uparrow \in P_F^\uparrow}
\sum_{k_1^\downarrow, \ldots, k_{k+1}^\downarrow \in P_F^\downarrow}
\idotsint 
\ud X_{[2,k+1]} \ud Y_{[2,k+1]}
\\ & \quad \times  
\left[
\prod_{j=1}^{k+1} 
e^{ik_j^\uparrow(x_j - x_{\pi(j)})}
e^{ik_j^\downarrow (y_j - y_{\tau(j)})}
\right]
\left[
\prod_{j=2}^{k+1}
g_s(x_j - y_j)
\right]
\\
& = 
\frac{1}{L^{6+6k}} 
\sum_{k_1^\uparrow, \ldots, k_{k+1}^\uparrow \in P_F^\uparrow}
\sum_{k_1^\downarrow, \ldots, k_{k+1}^\downarrow \in P_F^\downarrow}
e^{i\left(k_1^\uparrow - k_{\pi^{-1}(1)}^\uparrow\right) x_1}
e^{i\left(k_1^\downarrow - k_{\tau^{-1}(1)}^\downarrow\right)y_1}
\\ & \quad \times 
\prod_{j=2}^{k+1}
\int \ud x_j 
\left[ 
e^{i\left(k_j^\uparrow - k_{\pi^{-1}(j)}^\uparrow + k_j^\downarrow - k_{\tau^{-1}(j)}^\downarrow\right)x_j}
\int \ud y_j
\left( 
g_s(x_j - y_j) e^{-i\left(k_j^\downarrow - k_{\tau^{-1}(j)}^\downarrow\right) (x_j - y_j)}
\right)
\right]
\\
& = 
\frac{1}{L^{6 + 6k}}
\sum_{k_1^\uparrow, \ldots, k_{k+1}^\uparrow \in P_F^\uparrow}
\sum_{k_1^\downarrow, \ldots, k_{k+1}^\downarrow \in P_F^\downarrow}
e^{i\left(k_1^\uparrow - k_{\pi^{-1}(1)}^\uparrow\right) x_1}
e^{i\left(k_1^\downarrow - k_{\tau^{-1}(1)}^\downarrow\right)y_1}
\\ & \quad \times 
\left[
\prod_{j=2}^{k+1}
L^3
\chi_{\left(k_j^\uparrow - k_{\pi^{-1}(j)}^\uparrow = k_{\tau^{-1}(j)}^\downarrow - k_j^\downarrow\right)}
L^3
\hat g_s\left(k_j^\downarrow - k_{\tau^{-1}(j)}^\downarrow\right)
\right].
\end{aligned}
\]
Again, each factor $L^3 \hat g_s$ we may bound by $C a b^2$.
Moreover, since the diagram is linked we have for each $j$ that 
$\pi^{-1}(j)\ne j$ and/or $\tau^{-1}(j) \ne j$. 
(Otherwise the vertices $\{(j,\uparrow), (j,\downarrow)\}$
would be disconnected from the rest.)
Thus, each characteristic function is non-trivial, 
and hence effectively kills one of the $k^\sigma_j$-sums.
Each surviving $k^\sigma_j$-sum has at most $N_\sigma \leq N$ many summands.
Thus (uniformly in $x_1,y_1$)
\[
\abs{\Gamma_{D}^{1,1}} \leq \rho^2 (C a b^2 \rho)^k
\]
for any diagram $D$ of type (B2) with $k$ clusters of internal vertices, i.e. with $g$-graph as in \Cref{fig.B2.general}.
For any integer $K$ we have some finite $K$-dependent number of diagrams with $k\leq K$.
Concretely let $M_{k_0} < \infty$ be the number of type (B2) diagrams with $k=k_0$.
Thus, using \Cref{eqn.k.ng.diagrams} for diagrams with $k> K$, we get
\begin{equation}\label{eqn.sum.B2.diagrams.truncated}
\begin{aligned}
\sum_{\substack{p,q\geq 0 \\ p+q \geq 1}}
\frac{1}{p!q!} 
\abs{\sum_{\substack{D\in \mcL_{p,q}^{1,1} \\ D \textnormal{ of type (B2)}}}
\Gamma_D^{1,1}}
& \leq 
\sum_{\substack{k=1}}^{K}
\frac{1}{k!^2} \sum_{\substack{D\in \mcL_{k,k}^{1,1} \\ D \textnormal{ of type (B2)}}}
\abs{\Gamma_D^{1,1}}
+
\sum_{\substack{k=K+1}}^{\infty}
\frac{1}{k!^2} 
\abs{
\sum_{\substack{D\in \mcL_{k,k}^{1,1} \\ D \textnormal{ of type (B2)}}}
\Gamma_D^{1,1}}
\\
& \leq 
\sum_{k=1}^{K} \frac{M_k}{k!^2} \rho^2 (Cab^2\rho)^k
+ C \rho^{2} (C s a b^2 \rho (\log N)^3)^{K+1}
\\
& \leq C_Ka b^2 \rho^3 + C \rho^{2} (C s a b^2 \rho (\log N)^3)^{K+1}
\end{aligned}
\end{equation}
for some constant $C_K>0$ if $sab^2 \rho (\log N)^3$ is sufficiently small.
\end{proof}

\begin{remark}[{Upper bound on number of diagrams --- why we can't pick $K=\infty$}]
For an upper bound on the number of diagrams we first find an upper bound on the number of graphs.
All the underlying graphs look like \Cref{fig.B2.general}, but the labelling of the internal vertices may be different.
We are free to choose which white (internal) vertex connects to $(2,\uparrow)$ and so on. 
In total there are thus $q! = k!$ many possible graphs.

Next, to bound the number of diagrams with any given $g$-graph we may forget the constraint that the diagram has to be linked 
and consider all choices of $\pi\in \mcS_{k+1}$ and $\tau\in \mcS_{k+1}$ instead of just those, for which the diagram is linked.
For both $\pi$ and $\tau$ there are then $(k+1)!$ many choices. Thus for each graph $G$ 
there is at most $(k+1)!^2$ many linked diagrams of type (B2) with $g$-graph $G$.
Thus there are at most $k!(k+1)!^2$ diagrams of type (B2) with $k$ clusters of internal vertices.
With this bound the sum 
\[
  \sum_{k}
\frac{1}{k!^2} \sum_{\substack{D\in \mcL_{k,k}^{1,1} 
\textnormal{ of type (B2)}}}
\abs{\Gamma_D^{1,1}}
  \leq 
  \sum_k k! (k+1)^2 (Cab^2\rho)^k
\]
is not convergent.
This prevents us from taking $K=\infty$ in \Cref{eqn.sum.B2.diagrams.truncated} 
and using the exact calculations for all (infinitely many) diagrams of type (B2). 
\end{remark}

\begin{remark}[Higher spin]\label{rmk.higher.spin.no.diagrams}
For $S\geq 3$ values of the spin the evaluation of the diagrams is the same, 
but the combinatorics of counting how many diagrams there are for each given size is more complicated.
Still, there is only some finite $K$-dependent number of diagrams with $k(D)\leq K$ and thus 
(the appropriately modified version of)
\Cref{eqn.sum.B2.diagrams.truncated} is valid if $sab^2\rho (\log N)^3 < c_S$ for some constant $c_S > 0$.
\end{remark}

\subsubsection{\texorpdfstring{$(2,0)$- and $(0,2)$}{(2,0)- and (0,2)}-type terms}
We bound the term
\begin{equation}\label{eqn.type.(2.0).contribution}
    \iint \rho_{\textnormal{Jas}}^{(2,0)} \left[
    \abs{\frac{\nabla f_p(x_1 - x_2)}{f_p(x_1-x_2)}}^2
    + \frac{1}{2} v(x_1-x_2)\right]
    \ud x_1 \ud x_2.
\end{equation}
The term with $\rho_{\textnormal{Jas}}^{(0,2)}$ is completely analogous.
We may bound the $2$-particle density as follows.
\begin{lemma}\label{lem.(2.0).density}
There exist constants $c,C>0$ such that if $N_\uparrow = \# P_F^\uparrow > C$
and 
$s a b^2 \rho (\log N)^3 < c$,
then 
\[
\begin{aligned}
    \abs{\rho_{\textnormal{Jas}}^{(2,0)}}
    & \leq C f_p(x_1-x_2)^2 \rho^2 
    \left[a b^2 \rho + \rho^{2/3} |x_1-x_2|^2 \left[1 + sab^2 \rho (\log N)^4\right]\right].
\end{aligned}
\]
\end{lemma}
\noindent
This is essentially (a slightly modified version of) \cite[Lemma 4.1]{Lauritsen.Seiringer.2024}.
We give the proof at the end of this section.

Using now \Cref{lem.(2.0).density} we get 
\begin{equation}\label{eqn.contribution.final.(2.0).terms}
\begin{aligned}
\eqref{eqn.type.(2.0).contribution}
 & 
 \leq 
    CN \rho \int \left[|\nabla f_p|^2 + \frac{1}{2} f_p^2 v \right] 
    \left[ ab^2 \rho + \rho^{2/3} |x|^2 \left[1 + sab^2 \rho (\log N)^4\right]\right]
    \ud x
  \\ & 
  \leq C N a^2 b^2 \rho + C N \rho^{5/3} a^3 \left[1 + sab^2 \rho (\log N)^4\right]
\end{aligned}
\end{equation}
where we used that 
\[
  \int \left[|\nabla f_p|^2 + \frac{1}{2} f_p^2 v \right] |x|^2 \ud x \leq Ca_p^3 \leq C a^3,
  \qquad 
  \int \left[|\nabla f_p|^2 + \frac{1}{2} f_p^2 v \right] \ud x \leq Ca_p \leq C a.
\]
The first inequality follows directly from the definition of the scattering length, \Cref{def.scattering.length}.
The second inequality is a simple computation following from \Cref{lem.properties.scattering.function,eqn.f.scat}:
Using integration by parts and $f_{p0}(x) \geq 1 - a_p^3/|x|^3$ with equality outside the support of $v$ we have,
denoting the derivative in the radial direction by $\partial_r$,
\begin{equation*}
\begin{aligned}
  \int \left(\abs{\nabla f_p}^2 + \frac{1}{2}v f_p^2\right) \ud x
  & = 4\pi \int_0^b \left(\abs{\partial_r f_p}^2 r^{2} + f_p\partial_r^2 f_p r^{2} + 4 f_p \partial_r f_p r\right) \ud r 
  \\
  & = \frac{12\pi a_p^{3} /b^{2}}{1 - a_p^3/b^3} + 4\pi \left[b - 2 \int_{0}^b f_p^2 \ud r\right]
  \leq C a_p.
\end{aligned}
\end{equation*}
Finally, we give the

\begin{proof}[{Proof of \Cref{lem.(2.0).density}}]
\Cref{eqn.thm.gaudin.main} reads for $n=2,m=0$ (recall \Cref{eqn.rho1.translation.invariance})
\[
    \rho_{\textnormal{Jas}}^{(2,0)}
    = f_p(x_1-x_2)^2 
    \left[
    \rho_\uparrow^2
    +
    \sum_{\substack{p,q \geq 0}} \frac{1}{p!q!} \sum_{D\in \mcL_{p,q}^{2,0}} \Gamma_D^{2,0}\right].
\]
We split the diagrams into two types, according to whether $\nu^* = 0$ or $\nu^* \geq 1$
($\nu$ and $\nu^*$ are defined in \Cref{def.k.ng}).
We write 
\[
    \rho_\uparrow^2 
    + 
    \sum_{\substack{p,q \geq 0}} \frac{1}{p!q!} \sum_{D\in \mcL_{p,q}^{2,0}} \Gamma_D^{2,0}
    = \xi_0 + \xi_{\geq 1},
\]
where 
\[
    \xi_0 = \rho_\uparrow^2 + \sum_{\substack{p,q \geq 0}} \frac{1}{p!q!} 
    \sum_{\substack{D\in \mcL_{p,q}^{2,0} \\ \nu^*(D) = 0}} \Gamma_D^{2,0},
    \qquad 
    \xi_{\geq 1} = \sum_{\substack{p,q \geq 0}} \frac{1}{p!q!} 
    \sum_{\substack{D\in \mcL_{p,q}^{2,0} \\ \nu^*(D) \geq 1}} \Gamma_D^{2,0}.
\]
We will do a Taylor expansion of $\xi_0$ but not of $\xi_{\geq 1}$.
This is completely analogous to what is done in \cite[Proof of Lemma 4.1]{Lauritsen.Seiringer.2024}.
Consider first $\xi_{\geq 1}$.
By \Cref{thm.gaudin} and \Cref{eqn.k.ng.diagrams} we have $\xi_{\geq 1}\leq C a b^2 \rho^3$ 
uniformly in $x_1,x_2$ if $s a b^2 \rho (\log N)^3 < c$.

Consider next $\xi_{0}$. We do a Taylor expansion to second order around the diagonal.
For the zero'th order we have $\xi_0(x_1=x_2) + \xi_{\geq 1}(x_1 = x_2) = 0$ 
since $\rho_{\textnormal{Jas}}^{(2,0)}(x_1,x_2)$ vanishes for $x_1=x_2$.
The first order vanishes by the symmetry in $x_1$ and $x_2$.
Finally, we may bound the second derivatives $\partial^i_{x_1}\partial^j_{x_1} \xi_0$
by following the same procedure as in \cite[Proof of Lemma 4.1, Equations (4.15) to (4.20)]{Lauritsen.Seiringer.2024}.
This crucially uses the bounds in \Cref{eqn.derivative.lebesgue.constants}.
We give this argument for completeness.


Write (recalling \Cref{eqn.decompose.clusters.n+m.geq1} and using that the $k=0$ term together with 
$\rho_\uparrow^2$ give the two-particle density $\rho^{(2,0)}$ by Wick's rule)
\begin{multline*}
  \xi_0 = 
  \rho^{(2,0)} 
  + 
  \sum_{k=1}^\infty \frac{1}{k!} \sum_{\substack{n_1,\ldots,n_k\geq 0 \\ m_1,\ldots,m_k \geq 0 \\ \textnormal{For each } \ell: n_\ell + m_\ell \geq 2}}
  \frac{1}{\prod_{\ell} n_\ell! m_\ell !} \sum_{G_\ell \in \mcC_{n_\ell, m_\ell}} 
  \idotsint \ud X_{[3,2+\sum_\ell n_\ell]} \ud Y_{\sum_{\ell} m_\ell} 
  \left[\prod_{\ell=1}^k \prod_{e\in G_\ell} g_e\right] 
  \\ 
  \times 
  \left[
    \sum_{\substack{\pi \in \mcS_{2 + \sum_\ell n_\ell} \\ \tau \in \mcS_{\sum_\ell m_\ell}}}
    (-1)^{\pi} (-1)^\tau \chi_{( (\pi, \tau, \{(1,\uparrow)\}\cup \{(2,\uparrow)\} \cup \cup_\ell G_\ell) \textnormal{ linked})} 
    \prod_{i=1}^{2+\sum_\ell n_\ell} \gamma_{N_\uparrow}^{(1)}(x_i; x_{\pi(i)})
    \prod_{j=1}^{\sum_\ell m_\ell} \gamma_{N_\downarrow}^{(1)}(y_j; y_{\tau(j)})
  \right].
\end{multline*}
The only dependence on $x_1$ is in the $\gamma$-factors in $[\cdots]$.
Computing the second derivatives $\partial^i_{x_1}\partial^j_{x_1} \xi_0$
we see that they are sums of terms where one or two of the $\gamma$-factors gain the derivatives 
$\partial^i_{x_1}$ and $\partial^j_{x_1}$.
The term $[\cdots]$ above is the truncated correlation. So is its derivative
$\partial^i_{x_1}\partial^j_{x_1}[\cdots]$
now only some of the $\gamma$-factors carry derivatives.
To bound this term we do as in \Cref{sec.abs.conv} and use the (appropriately modfied) formula in \Cref{eqn.BBF.rho_t}.
The $\gamma$-factors with derivative can either end up in the anchored tree, or in the matrix $\mcR(r)$.
Following the argument in \Cref{sec.Gamma.nm.abs.conv} to bound $\partial^i_{x_1}\partial^j_{x_1} \xi_0$
we see that we need bounds on the determinants of the matrix $\mcR(r)$, modified with the $\gamma$-factors with derivatives,
and/or of the integrals of $\gamma$-factors with derivatives.

If the $\gamma$-factors with derivatives end up in the matrix $\mcR(r)$ we gain a factor $C\rho^{1/3}$ in the bound of 
its determinant, \Cref{eqn.bound.det.N(r)}. 
This follows from a slight modification of \Cref{eqn.bound.det.N(r)} 
and is explained around \cite[Equation (D.9)]{Giuliani.Mastropietro.ea.2021}:
One changes the definition of some of the functions $\alpha_\mu$ in the proof of \Cref{eqn.bound.det.N(r)}
by including factors $ik^i$ and/or $ik^j$.
If the $\gamma$-factors with derivatives end up in the anchored tree 
we either have to bound them pointwise, in which case we gain a factor $C\rho^{1/3}$,
or we have to bound their integrals, in which case we use \Cref{eqn.derivative.lebesgue.constants}.

Following the argument in \Cref{sec.Gamma.nm.abs.conv} we thus get a bound similar to \Cref{eqn.final.bound.Gamma.nm}
with the following modifications: 
One or two factors $\int_{\Lambda} \abs{\gamma^{(1)}_{N_\uparrow}}$
is replaced with factors with derivatives 
$\int_{\Lambda} \abs{\partial_1 \gamma^{(1)}_{N_\uparrow}}, \int_{\Lambda} \abs{\partial_2 \gamma^{(1)}_{N_\uparrow}}$,
where $\partial_{1}, \partial_2 \in \{1, \partial^i_{x_1}, \partial^j_{x_1}, \partial^{i}_{x_1}\partial^{j}_{x_1}\}$
are the derivatives hitting $\gamma$-factors in the anchored tree that we do not bound pointwise. 
For diagrams with only one internal cluster (i.e. with $k=1$) there is only one such $\gamma$-factor.
Moreover we gain a factor $C\rho^{(2-\#\partial_1 - \#\partial_2)/3}$ 
where $\#\partial_j$ denotes the number of derivatives in $\partial_j$,
i.e. $\# 1 = 0, \# \partial^i_{x_1} = 1$ and $\# \partial^{i}_{x_1}\partial^{j}_{x_1} = 2$.
This factor arises from the matrix $\mcR(r)$, modified to include the derivatives, 
and the $\gamma$-factors with derivatives we bound pointwise.
The derivatives in either (the modification of) 
$\mcR(r)$ or on $\gamma$-factors we bound pointwise are exactly those not in $\partial_1$ or $\partial_2$.
That is,

\[
\begin{aligned}
  \abs{\partial^i_{x_1}\partial^j_{x_1} \xi_0}
  & \leq 
  \abs{\partial^i_{x_1}\partial^j_{x_1} \rho^{(2,0)}}
  +
  C \rho^{2} 
  \vast[
  \sum_{\partial \in \{1, \partial^i_{x_1}, \partial^j_{x_1}, \partial^{i}_{x_1}\partial^{j}_{x_1}\}}
  \rho^{(2-\#\partial)/3} \int_\Lambda \abs{\partial \gamma_{N_\uparrow}^{(1)}}
  \sum_{\substack{n_0, m_0 \geq 0 \\ n_0 + m_0 \geq 2}}
      (C a b^2 \rho)^{n_0 + m_0 - 1}
  \\ & \qquad 
  +
  \sum_{ \substack{\partial_1, \partial_2 \in \{1, \partial^i_{x_1}, \partial^j_{x_1}, \partial^{i}_{x_1}\partial^{j}_{x_1}\}
  \\ \partial_1\partial_2 \in \{1, \partial^i_{x_1}, \partial^j_{x_1}, \partial^{i}_{x_1}\partial^{j}_{x_1}\}}}
  \rho^{(2-\#\partial_1 - \#\partial_2)/3} 
  \int_\Lambda \abs{\partial_1 \gamma_{N_\uparrow}^{(1)}}
  \int_\Lambda \abs{\partial_2 \gamma_{N_\uparrow}^{(1)}}
  \\ & \qquad \quad \times 
  \sum_{k=2}^\infty 
    \left[
    C s (\log N)^3
      \right]^{k-1}
      \left[
    \sum_{\substack{n_0, m_0 \geq 0 \\ n_0 + m_0 \geq 2}}
      (C a b^2 \rho)^{n_0 + m_0 - 1}
      \right]^k
    \vast].
\end{aligned}
\]
Noting that $\abs{\partial_{x_1}^i\partial_{x_1}^{j} \rho^{(2,0)}} \leq C \rho^{8/3}$
by a simple computation using the Wick rule and 
using \Cref{eqn.derivative.lebesgue.constants} to bound the integrals 
we conclude that
\[
\begin{aligned}
\abs{\partial^i_{x_1}\partial^j_{x_1} \xi_0}
& \leq C \rho^{8/3} \left[1 + sab^2 \rho (\log N)^4\right]
\end{aligned}
\]
if $N_\uparrow$ is sufficiently large and $s a b^2 \rho (\log N)^3$ is sufficiently small.
By Taylor's theorem we conclude the desired.
\end{proof}

\subsection{\texorpdfstring{$3$}{3}-body terms}
In this section we bound the $3$-body terms of \Cref{eqn.energy.compute}.

\subsubsection{\texorpdfstring{$(2,1)$- and $(1,2)$}{(2,1)- and (1,2)}-type terms}
We bound the term 
\begin{equation}\label{eqn.type.(2.1).terms}
    \iiint \rho_{\textnormal{Jas}}^{(2,1)} 
    \left[
    \abs{\frac{\nabla f_s(x_1-y_1)\nabla f_p(x_1-x_2)}{f_s(x_1-y_1)f_p(x_1-x_2)}} 
    + \abs{\frac{\nabla f_s(x_1-y_1) \nabla f_s(x_2-y_1)}{f_s(x_1-y_1) f_s(x_2-y_1}}
    \right]
    \ud x_1 \ud x_2 \ud y_1.
\end{equation}
The $(1,2)$-type term is bounded analogously.

By \Cref{thm.gaudin} we have the bound 
\[
\rho_{\textnormal{Jas}}^{(2,1)} \leq C \rho^3 f_s(x_1-y_1)^2 f_s(x_2-y_1)^2 f_p(x_1- x_2)^2
\]
if $s a b^2 \rho (\log N)^3$ is sufficiently small.
In the first summand in \Cref{eqn.type.(2.1).terms} we moreover bound $f_s(x_2-y_1) \leq 1$ and in the second summand we bound $f_p(x_1-x_2) \leq 1$.
Then by the translation invariance we have
\[
\eqref{eqn.type.(2.1).terms}
\leq 
C N \rho^2 
    \left[ 
    \left(\int f_s |\nabla f_s|\right) \left(\int f_p |\nabla f_p|\right)
 + \left(\int f_s |\nabla f_s|\right)^2
    \right].
\]
By radiality and \Cref{lem.properties.scattering.function} we have 
\begin{multline*}
\frac{1}{4\pi}\int f_s |\nabla f_s|
 = \int_0^b r^2 f_s \partial_r f_s \ud r
 = \frac{1}{2}[r^2 f_s^2]_0^b 
    - \frac{1}{2} \int_0^b 2r f_s^2 \ud r
\\
\leq \frac{1}{2}b^2 - \frac{1}{(1 - a/b)^2}\int_a^b r \left(1 - \frac{a}{r}\right)^2 \ud r 
\leq C ab,
\end{multline*}
where $\partial_r$ denotes the radial derivative.
Similarly by \Cref{lem.properties.scattering.function}
\begin{multline*}
\frac{1}{4\pi}\int f_p |\nabla f_p|
 = \int_0^b r^2 f_p \partial_r f_p \ud r
 = \frac{1}{2}[r^2 f_p^2]_0^b 
    - \frac{1}{2} \int_0^b 2r f_p^2 \ud r
\\
\leq \frac{1}{2}b^2 - \frac{1}{(1 - a_p^3/b^3)^2}\int_a^b r \left(1 - \frac{a_p^3}{r^3}\right)^2 \ud r 
\leq C a_p^2.
\end{multline*}
We conclude that (for sufficiently small $s a b^2 \rho (\log N)^3$)
\begin{equation}\label{eqn.contribution.final.(2.1).terms}
\eqref{eqn.type.(2.1).terms}
\leq C N \rho^2 a^2 b^2.
\end{equation}

\subsubsection{\texorpdfstring{$(3,0)$- and $(0,3)$}{(3,0)- and (0,3)}-type terms}
We may bound 
\begin{equation}\label{eqn.contribution.final.(3.0).terms}
\iiint \rho_{\textnormal{Jas}}^{(3,0)} 
\abs{ \frac{\nabla f_p(x_1-x_2) \nabla f_p(x_1-x_3)}{f_p(x_1-x_2)f_p(x_1-x_3)}}
\ud x_1 \ud x_2 \ud x_3
\leq C N \rho^2 a^4
\end{equation}
using the same method as for the $(2,1)$-type terms.
The $(0,3)$-type terms may be bounded analogously.

\begin{remark}[Higher spin]\label{rmk.higher.spin.(1.1.1)}
For higher spin we also have terms of type $(1,1,1)$. 
These may be bounded exactly as the $(2,1)$-type terms with 
two $s$-wave factors.
\end{remark}

\subsection{Putting the bounds together}
Combining \Cref{eqn.energy.compute,eqn.contribution.final.(2.0).terms,eqn.contribution.final.(2.1).terms,eqn.contribution.final.(3.0).terms,eqn.contribution.final.(1.1).terms,eqn.energy.fermi.polyhedron}
we immediately get for any integer $K$
\begin{equation}
\begin{aligned}
& 
\frac{\longip{\psi_{N_\uparrow,N_\downarrow}}{H_N}{\psi_{N_\uparrow,N_\downarrow}}}{L^3}
\\ & \quad  = 
\frac{3}{5} (6\pi^2)^{2/3} \left(\rho_\uparrow^{5/3} + \rho_{\downarrow}^{5/3}\right) 
+ 8\pi a \rho_\uparrow\rho_\downarrow
+ O((s_\uparrow^{-2} + s_\downarrow^{-2})\rho^{5/3}) + O(N^{-1/3} \rho^{5/3})
\\ & \qquad 
+ O(a^2 b^{-1} \rho^2) 
+ O_K(a^2b^2 \rho^3)
+ O_K(a\rho^2 (sab^2\rho (\log N)^3)^{K+1})
+ O(s a^3 \rho^3 \log(b/a) (\log N)^3)
\\ & \qquad 
+ O(a^2 b^2 \rho^3)
+ O\left(\rho^{8/3}a^3\left[1 + sab^2\rho(\log N)^4\right]\right)
\end{aligned}
\end{equation}
for $s a b^2 \rho (\log N)^3$ sufficiently small and $N_\sigma = \# P_F^\sigma$ sufficiently large.
As in \cite[Section 4]{Lauritsen.Seiringer.2024} we will choose $N_\sigma$ some large negative power of $a^3\rho$.
By choosing, say, $L \sim a(a^3\rho)^{-10}$ (still requiring that $\frac{k_F^\sigma L}{2\pi}$ is rational) 
we have $N \sim (a^3\rho)^{-29}$.
(More precisely one chooses $L \sim a ((k_F^\uparrow + k_F^\downarrow) a)^{-30}$, see \Cref{rmk.dependent.parameters}.)
Additionally we choose 
\[
  s_\sigma \sim (a^3\rho)^{-1/3 + \eps},
\]
where $\eps > 0$ is chosen as $\eps = \frac{1}{K}$ for $K>6$. Recall moreover that $b=\rho^{-1/3}$.
Thus for any fixed integer $K > 6$ we have 
\begin{equation}\label{eqn.energy.in.box}
\frac{\longip{\psi_{N_\uparrow,N_\downarrow}}{H_N}{\psi_{N_\uparrow,N_\downarrow}}}{L^3}
= 
\frac{3}{5} (6\pi^2)^{2/3} \left(\rho_\uparrow^{5/3} + \rho_{\downarrow}^{5/3}\right) 
+ 8\pi a \rho_\uparrow\rho_\downarrow
+ O_K\left(a\rho^{2} (a^3\rho)^{1/3-2/K}\right).
\end{equation}

\subsection{Box method}\label{sec.box}
We extend to the thermodynamic limit using a box method exactly as in \cite[Section 4.1]{Lauritsen.Seiringer.2024}. We sketch the details here.
Using a bound of Robinson \cite[Lemmas 2.1.12, 2.1.13]{Robinson.1971} (more specifically the form in \cite[Section C]{Mayer.Seiringer.2020}, see also \cite[Lemma 4.3]{Lauritsen.Seiringer.2024})
we have an isometry $U$ such that $U\psi_{N_\uparrow,{N_\downarrow}}$ 
has Dirichlet boundary conditions in the box $\Lambda_{L+2d} = [-L/2-d, L/2+d]^3$ and
\[
\longip{U\psi_{N_\uparrow,N_\downarrow}}{H_{N,L+2d}^{D}}{U\psi_{N_\uparrow,{N_\downarrow}}}
\leq \longip{\psi_{N_\uparrow,{N_\downarrow}}}{H_{N,L}^{\textnormal{per}}}{\psi_{N_\uparrow,{N_\downarrow}}}
+ \frac{6N}{d^2},
\]
where $H_{N,L+2d}^{D}$ denotes the Hamiltonian on a box of sides $L+2d$ with Dirichlet boundary conditions, 
and $H_{N,L}^{\textnormal{per}}$ denotes the Hamiltonian on a box of sides $L$ with periodic boundary conditions.
We are free to choose the the parameter $d$. We will choose it some large negative power of $a^3\rho$.

We use this to form trial states $U\psi_{N_\uparrow,N_\downarrow}$ with Dirichlet boundary conditions
in a box of sides $L + 2d$.
Using then a box method of glueing copies of the trial state $U\psi_{N_\uparrow,N_\downarrow}$ together (as in \cite[Section 4.1]{Lauritsen.Seiringer.2024})
with a distance $b$ between them (same $b$ as before)
we get a trial state $\Psi_{M^3 N_\uparrow, M^3 N_\downarrow}$ of particle densities 
$\tilde \rho_\sigma = \frac{M^3N_\sigma}{M^3 (L + 2d + b)^3} = \rho_\sigma (1 + O(b/L) + O(d/L))$.
The state $\Psi_{M^3 N_\uparrow, M^3 N_\downarrow}$ has the energy density
\[
\begin{aligned}
  & \frac{\longip{\Psi_{M^3 N_\uparrow, M^3 N_\downarrow}}{H_{M^3 N,M^3(L+2d+b)}^D}{\Psi_{M^3 N_\uparrow, M^3 N_\downarrow}}}{M^3(L + 2d + b)^3}
  \\ & \quad 
  = \frac{\longip{U\psi_{N_\uparrow,N_\downarrow}}{H_{N,L+2d}^{D}}{U\psi_{N_\uparrow,{N_\downarrow}}}}{L^3} \left(1 + O(d/L) + O(b/L)\right)
  \\ & \quad 
  \leq 
  \frac{\longip{\psi_{N_\uparrow,{N_\downarrow}}}{H_{N,L}^{\textnormal{per}}}{\psi_{N_\uparrow,{N_\downarrow}}}}{L^3}\left(1 + O(d/L) + O(b/L)\right)
  + O(\rho d^{-2}).
\end{aligned}
\]
Choosing say $d = a(a^3\rho)^{-5}$ and using \Cref{eqn.energy.in.box} we thus get 
\[
\begin{aligned}
  e(\tilde\rho_\uparrow,\tilde\rho_\downarrow)
  & \leq \limsup_{M\to \infty} 
  \frac{\longip{\Psi_{M^3 N_\uparrow, M^3 N_\downarrow}}{H_{M^3 N,M^3(L+2d+b)}^D}{\Psi_{M^3 N_\uparrow, M^3 N_\downarrow}}}{M^3(L + 2d + b)^3}
  \\ & \leq 
\frac{3}{5} (6\pi^2)^{2/3} (\rho_\uparrow^{5/3} + \rho_\downarrow^{5/3})
+ 8\pi a \rho_\uparrow\rho_\downarrow 
+ O_K\left(a\rho^2 (a^3\rho)^{1/3-2/K}\right)
  \\
  & = 
  \frac{3}{5} (6\pi^2)^{2/3} (\tilde\rho_\uparrow^{5/3} + \tilde\rho_\downarrow^{5/3})
+ 8\pi a \tilde\rho_\uparrow\tilde\rho_\downarrow 
+ O_K\left(a\tilde\rho^2(a^3\tilde \rho)^{1/3-2/K}\right)
\end{aligned}
\]
since $\tilde \rho_\sigma = \rho_\sigma(1 + O( (a^3\rho)^{-5}))$.
For any $\delta > 0$ we may take $K > (2\delta)^{-1}$. This concludes the proof of \Cref{thm.main} 
for pairs of densities $(\tilde \rho_\uparrow, \tilde\rho_\downarrow)$ arising from the construction above. 
As noted in \Cref{rmk.dependent.parameters} this is not all possible values of the densities $\rho_\sigma$.
Finally, we extend \Cref{thm.main} to all pairs of (sufficiently small) densities.

Consider any pair of densities $(\rho_{\uparrow0},\rho_{\downarrow0})$ and define $\rho_0 = \rho_{\uparrow0} + \rho_{\downarrow 0}$
and the Fermi momenta $k_{F0}^{\sigma} := (6\pi^2)^{1/3} \rho_{\sigma 0}^{1/3}$.
Let $\eps > 0$ be some small parameter to be chosen later and find (by density of the rationals in the reals) $k_F^\sigma$ with 
$(1 + \eps) k_{F0}^\sigma \leq k_F^\sigma \leq (1 + 2\eps) k_{F0}^\sigma$ and $k_F^\uparrow / k_F^{\downarrow}$ rational (recall \Cref{rmk.k_F.rational}).
Following the construction above we find a trial state $\psi_{N_\uparrow, N_\downarrow}$ with particle densities $\rho_\sigma$
satisfying 
\[
(1 + 3\eps + O(\eps^2) + O(N_\sigma^{-1/3})) \rho_{\sigma 0} 
\leq \rho_\sigma 
\leq (1 + 6\eps + O(\eps^2) + O(N_\sigma^{-1/3})) \rho_{\sigma 0}.
\]
Thus by constructing the trial states $\Psi_{M^3N_\uparrow, M^3N_\downarrow}$ of particle densities $\tilde\rho_\sigma$ as above 
we find
\[ 
\left(1 + 3\eps + O(\eps^2) +  O( (a^3\rho_0)^{-5})\right) \rho_{\sigma 0} 
\leq \tilde\rho_\sigma 
\leq \left(1 + 6\eps + O(\eps^2) + O( (a^3\rho_0)^{-5})\right) \rho_{\sigma 0}.
\]
Choosing then $\eps = (a^3\rho_0)^{-4}$ we have $\rho_{\sigma 0} \leq \tilde \rho_\sigma$
and $\tilde \rho_\sigma = \rho_{\sigma 0} (1 + O((a^3\rho_0)^{-4}))$
for sufficiently small $a^3\rho_0$.
Since $v\geq 0$ the energy is monotone increasing in the particle number, thus so is the energy density.
Hence for any $\delta > 0$
\[
\begin{aligned}
  e(\rho_{\uparrow 0}, \rho_{\downarrow 0}) 
  & \leq e(\tilde\rho_\uparrow,\tilde\rho_\downarrow)
  \\
  & \leq 
  \frac{3}{5} (6\pi^2)^{2/3} (\tilde\rho_\uparrow^{5/3} + \tilde\rho_\downarrow^{5/3})
+ 8\pi a \tilde\rho_\uparrow\tilde\rho_\downarrow 
+ O_\delta(a\rho^2 (a^3\tilde\rho)^{1/3-\delta})
  \\
  & = 
  \frac{3}{5} (6\pi^2)^{2/3} (\rho_{\uparrow0}^{5/3} + \rho_{\downarrow0}^{5/3})
+ 8\pi a \rho_{\uparrow0}\rho_{\downarrow0} 
+ O_\delta(a\rho_0^2 (a^3\rho_0)^{1/3-\delta}).
\end{aligned}
\]
This concludes the proof of \Cref{thm.main}.

\subsubsection*{Acknowledgements}
We thank Alessandro Giuliani and Robert Seiringer for helpful discussions and Robert Seiringer for his comments on the manuscript.
Financial support by the Austrian Science Fund (FWF) through 
grant \href{https://doi.org/10.55776/I6427}{DOI: \nolinkurl{10.55776/I6427}}
(as part of the SFB/TRR~352) is gratefully acknowledged.

\printbibliography
\end{document}